\documentclass[manuscript,onecolumn]{geophysics}
\usepackage{graphicx}
\usepackage{subfigure}
\usepackage[english]{babel}
\usepackage{amsmath, amsthm, amssymb}
\usepackage{mathtools}
\usepackage{geometry}
\usepackage{setspace}
\usepackage[dvipsnames]{xcolor}
\usepackage{rotating}

\newcommand{\defeq}{\mathrel{\mathop:}=}
\newcommand  {\x}{{\bf x}}

\begin{document}
\title{Overview of Physics-Informed Machine Learning Inversion
of Geophysical Data}

\renewcommand{\thefootnote}{\fnsymbol{footnote}}
\address{
\footnotemark[1]Department of Earth Science and Engineering, King Abdullah University of Science and Technology \\
\footnotemark[2]Theoretical Division, Los Alamos National Laboratory \\
}
\author{Gerard T. Schuster\footnotemark[1] and Shihang Feng\footnotemark[2]}

\footer{Manuscript}
\lefthead{Schuster&Shihang}
\righthead{Physics-Informed, Inversion, Machine Learning}

\maketitle
\begin{center}
{\LARGE
{\bf Overview of Physics-Informed Machine Learning Inversion
of Geophysical Data}}
\mbox{}\\
\mbox{}\\
{\Large \bf
Gerard T. Schuster and Shihang Feng}
\mbox{}\\
\mbox{}\\
{\Large \bf ABSTRACT}
\end{center}
{\it The second-best way to solve a problem is by knowing the physics that underlies the problem. The best way might be to understand the physics and use a strategy of physics-based machine learning.}
\mbox\\
\mbox{}\\
\mbox{}\\
We review four types of
algorithms for physics-informed machine learning (PIML) inversion
of geophysical data.
The unifying equation is given by
the joint objective function $\epsilon$:
\begin{eqnarray}
\epsilon^{||-PIML}&=&\lambda_1 \overbrace{||{\bf W}^{ML}({\bf H}_{{\bf w}} {\bf d}^{obs}-{\bf m})||^2}^{NN}
+ \lambda_2 \overbrace{{||{\bf W}^{FWI}({\bf L} {\bf m}-{\bf d}^{obs})||^2}}^{FWI} ~+
\nonumber\\
\nonumber\\
&&
+ ~~Regularizer,
\label{PIML.eq120}
\end{eqnarray}
where the optimal model ${\bf m}^*$ and weights $\bf w^*$
minimize $\epsilon$.
Here, The matrix weights
are given by the boldface symbol $\bf W$, and
full waveform inversion (FWI) is typically computed using
a finite-difference solution of the wave equation, where $\bf L$ represents the forward modeling operation of the
wave equation as a function of the model $\bf m$. Also,
a fully-connected neural network (NN) is used to compute the model ${\bf H_w}{\bf d}^{obs} \approx \bf m$ from the
observed input
data ${\bf d}^{obs}$. The selection of weights $\lambda_i$ and
the NN operations determine one of four different
PIML algorithms.

PIML offers potential advantages over standard FWI through its enhanced ability to avoid local minima and the option to locally train the inversion operator, minimizing the requirement for extensive training data for global applicability. However, the effectiveness of PIML relies on the similarity between the test and trained data. Nevertheless, a possible strategy to overcome this limitation involves initial pretraining of a PIML architecture with data from a broader region, followed by fine-tuning for specific data—a method reminiscent of the way large language models are pretrained and adapted for various tasks.

\section{Introduction}

There are many examples
of geophysical inversion, including full waveform
inversion (FWI) of the velocity model $\mathbf{m}$ from
seismic data $\mathbf{d}$~\citep{virieux2009overview}
or estimating the resistivity model from transient electromagnetic (TEM)  data\footnote{We will often refer to the inverted model as a velocity model,
but it is to be understood that it can also be a resistivity model in the context
of TEM,  a density model with respect to gravity inversion and so on.}.
These procedures use a
physics-based modeling method with a gradient-descent algorithm,
which requires expensive numerical solutions to the governing
equations, such as the seismic wave equation or Maxwell's equations.
For FWI, the optimal velocity model $\mathbf{m}^*$ minimizes the regularized
sum of the squared residuals $\epsilon^{FWI}$ in equation~\ref{PIML.eq120}
for $\lambda_1=0$ and $\lambda_2=1$
to get:
 \begin{eqnarray}
 \mathbf{m}^*&=& argmin_{\mathbf{m}} [~ \overbrace{
 ||{ \bf W^{FWI}} (\mathbf{L} \mathbf{m}-\mathbf{d})||^2 ~~+ ~Regularizer
 }^{\epsilon^{FWI}}
 ~],
 \label{ch.InvIntro.eq1aa}
 \end{eqnarray}
 where the regularizer is designed, for example, to favor desirable models and ${\bf  W}^{FWI}$ is a matrix that downweights
 unreliable parts of the data.
 For non-linear problems such as FWI\index{Full waveform inversion (FWI)}, the forward modeling operator $\mathbf{L}$ is a forward modeling
  operator associated with the velocity model $\mathbf{m}$, e.g. a finite-difference
 solution to the wave equation,
 and the optimal model $\mathbf{m}^*$ is typically found by a non-linear gradient-descent algorithm.

\begin{figure}
\includegraphics[width=1\columnwidth]{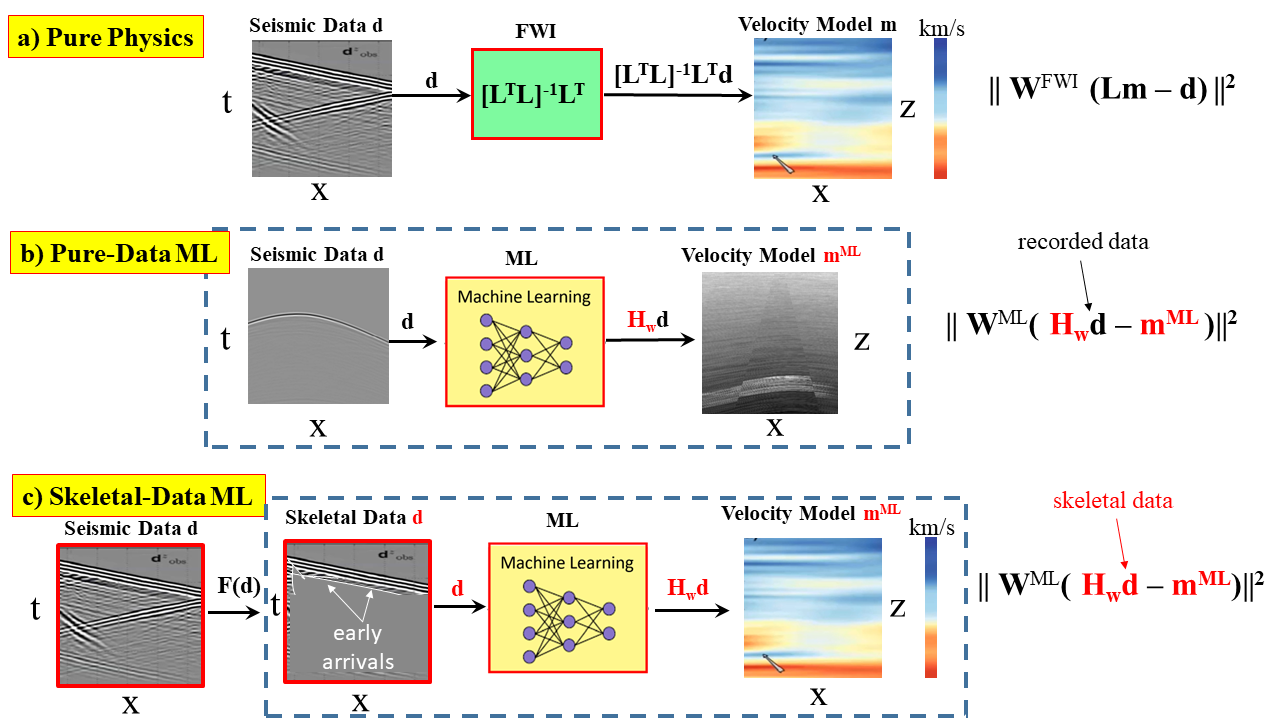}
\caption{Schematics for data inversion by a) pure-physics FWI, b) pure-data ML, and
c) skeletal-data ML where $F({\bf d})$ is a filtering operation
that extracts skeletal data from the recorded data. The dashed rectangles
indicate that the enclosed system trains
on many examples to get the optimal parameters $\mathbf{w}$
that minimize $||{\bf H}_{{\bf w}} {{\bf d}} - {\bf m}^{ML}||^2$.
The objective functions for each algorithm are on the right.}
\label{ch.ML.Inv1}
\end{figure}

\begin{figure}
\includegraphics[width=1\columnwidth]{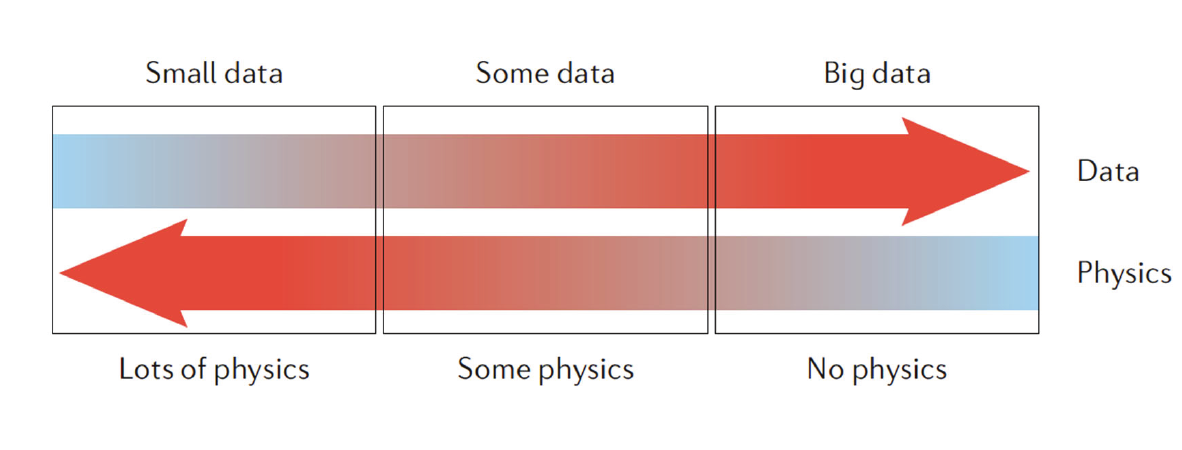}
\caption{Diagram illustrating
 physics-related constraints as
a function of data size~\citep{karniadakis2021physics}. Here, physics provides extra
information that can make up for the lack of sufficient data in the training of an ML algorithm.}
\label{ch.ML.Inv4}
\end{figure}

 The computational schematic\index{Workflow: FWI} for the general FWI inversion
 algorithm is illustrated in Figure~\ref{ch.ML.Inv1}a.
 This deterministic algorithm is completely physics based and  is typically applied to seismic data, as represented on the
far left of the spectrum in Figure~\ref{ch.ML.Inv4}. Here,  no data other than the shot gathers
are typically used for inversion.

In contrast to FWI, the prediction of a model by a
trained neural network (NN)
is completely
devoid of the explicit constraints of wave-propagation  physics. It is an inversion in the sense that
the NN parameters ${\bf w}$ are obtained by training to
predict the given model
${\bf m}={\bf H}_{{\bf w}} {\bf d}$ from the training
pairs $({\bf m},{\bf d})\in training~data$.
Here, ${\bf H}_{{\bf w}}$ represents the non-linear operator of the NN, which is a function of the nodal parameters denoted by the vector ${\bf w}$.
The ML goal is to find the optimal parameters ${\bf w}^*$
 that minimize the NN objective function in equation~\ref{PIML.eq120}
 for $\lambda_2=0$ (see
Figure~\ref{ch.ML.Inv1}b) s.t.
\begin{eqnarray}
{\bf w}^* &=& argmin_{{\bf w}} [~\overbrace{
||{\bf W}^{ML}(
{\bf H}_{{\bf w}} {\bf d}-{\bf m})||^2 ~~ +~ Regularizer}^{\epsilon^{ML}}~],
 \label{ch.InvIntro.eq4}
\end{eqnarray}
which is typically achieved by a gradient-descent algorithm.
Here, ${\bf W}^{ML}$ is the matrix used to optimize
convergence such as balancing the number of input classes~\citep{shi2023semi}. Once the network is accurately trained for a generalized set of training pairs, then ${\bf H}_{{\bf w}} $
can be applied to new data to get the associated model.

If the input traces are skeletonized by muting everything but first arrivals
then Figure~\ref{ch.ML.Inv1}c depicts the schematic for skeletal-data ML.
Some successful examples
that use skeletonized data
for training including the inversion of early arrivals
in seismic data~\citep{yu2021skeletonized}, recognition of
 different bird types in pictures of birds~\cite{mao2021domain} and rock cracks in images of a canyon wall~\citep{shi2023semi}.
The case of pure-data NN is represented on the
far right of the spectrum in Figure~\ref{ch.ML.Inv4}, where a huge number
of
diverse training pairs must be used to accurately estimate the ML model.

The problem with pure-data ML inversion
is that it takes an enormous number of training examples to thoroughly train the NN
to invert arbitrary test data.
Incomplete training, where the training set is limited in its diversity and number of examples, is inevitable for high-dimensional data spaces:
increasing the dimension of the data space leads to a shrinkage of
the distance between the nearest and farthest data points~\citep{durrant2009nearest}.
This shrinkage can be fatal for inversion methods that
use a limited number of diverse training examples. Thus, achieving a
marginal improvement in generalizing an NN \index{Neural network (NN)} requires an exponential increase in the number of training examples. As~\cite{bishop2006pattern} states: {\it increasing the input space dimension without enhancing the quantity of available information
reduces the model's power and may give rise to the curse of dimensionality.}
Fortunately, the problem with incomplete data can be mitigated by using
joint inversion to include
the constraints of physics, which is a proxy for using new data.

\subsection{Joint inversion}
Joint geophysical inversion is an approach for simultaneously inverting different types of geophysical data~\citep{vozoff1975joint,gallardo2004joint,abubakar2012joint,zhdanov2021advanced}, where the
objective function is the summation of regularization terms
and misfit functions for each type of data. Typically, the different data sets are recorded by different instruments such as a gravimeter that
records the gravity data and geophones that record the seismic data. The density
model reconstructed from the gravity data can have structural characteristics similar to the velocity tomogram reconstructed from the seismic data,
so a regularization term is added to the misfit function that encourages
structural similarities between the gravity and seismic models.
As an example,~\cite{paulatto2019vertically} utilized a joint interpretation of seismic and gravity data for characterizing a
magma reservoir.
Other examples include the inversion of resistivity and traveltime
data by~\cite{gallardo2004joint} to image the near surface, and a more comprehensive
inversion is by~\cite{colombo2007geophysical} who jointly inverted seismic, gravity, and electromagnetic data for velocity
modeling.~\cite{juhojuntti2015joint} inverted both seismic refraction and resistivity data for imaging the subsurface properties
associated with
groundwater.

If both a statistical ML method and a physics-based FWI are used to compute
a velocity model, then these two methods can be considered different
from one another but their final models should agree with one another.
To encourage this agreement, their corresponding misfit functions can
be added together along with a regularization term to give the
 joint objective function in equation~\ref{PIML.eq120}.
This joint objective function gives the unifying approach used by physics-informed machine learning (PIML)
 inversion
 of geophysical data.

\subsection{PIML}
To partially overcome the problem of incomplete ML training,~\cite{colombo2021framework} computed the optimal model $\bf m^*$ and weights $\bf w^*$ in equation~\ref{PIML.eq120}
that minimized
\begin{eqnarray}
({\mathbf w}^*,{\bf m}^*)&=&argmin_{{\mathbf w},{\mathbf m} }
[\lambda_1 ||{\bf W}^{ML}({\bf H}_{{\bf w}} {\bf d}^{obs}-{\bf m}^{ML})||^2
+ \lambda_2 {||{\bf W}^{FWI}({\bf L} {\bf m}-{{\bf d}}^{obs})||^2} +\nonumber\\
\nonumber\\
&&\lambda_3||{\bf W}^{joint} [{\bf m}-{\bf m}^{ML}]||^2
+ ~~ Regularizer],
\label{PIML.eq1}
\end{eqnarray}
where $||{\bf W}^{joint} [{\bf m}-{\bf m}^{ML}]||^2$ is
a regularizer that {\it softly}
constrains the FWI and NN models to agree with one another.
The weighting matrices ${\bf W}^{ML}$, ${\bf W}^{FWI}$ and ${\bf W}^{joint}$ are selected to
optimize convergence and account for noisy data.
Here, ${\bf d}^{obs}$ represents the observed data, $\lambda_i ~i \in [1, 2, 3]$ are regularization parameters and
the optimal solution ${\bf m}^*$ can be found, for example, by
a gradient-descent method that alternates between solving for the
velocity model $\bf m$ and the ML weights ${\bf w}$.
Table 1 summarizes four different PIML algorithms which are determined by
 the
 values of $\lambda_i$
 and the selection of the deterministic and NN schemes.

\begin{table}[!ht]
{\scriptsize
    \centering
\begin{minipage}[t]{1\linewidth}
\caption{\footnotesize PIML algorithms from equation~\ref{PIML.eq1}, where $(N_{AE},N_{FLI},N_{FWI},N_{it})$ are the number of iterations
for the $(autoencoder, ~FLI, ~FWI,~ FWI+ML)$ schemes. The  ${\bf L m}$ represents
the finite-difference solution to the wave equation for the velocity model $\bf m$ and ${\bf d^{pred}}=
{\bf L_v}({\bf x},{\bf m^{ML}}) $ is the NN solution to the wave equation
for the ML velocity model ${\bf m}^{ML}$.
The joint regularization function
is given by $\epsilon^{const.}=\lambda_3||{\bf m}-{\bf m}^{ML}||^2$
and the weight matrices have been conveniently neglected.}
\label{PIML.Tab1}
\begin{tabular}{|c |c |c|} 
\hline 
&&\\
{\bf PIML Type}  &  {\bf Schedule}& {\bf NN Architecture} \\   \hline
    &&\\
Skeletal& $(\lambda_1,\lambda_2,\lambda_3)=(1,0,0)$ & ${\bf H_w}$ = Autoencoder
    \\
PIML&$for~ i=1:N_{AE}$&\\
 &
$(\lambda_1,\lambda_2,\lambda_3)=(0,1,0)$
&  $\epsilon^{ML}=\lambda_1||{\bf H_w} {\bf d^{obs}}- {\bf d^{obs}}||^2$;~
$\epsilon^{FLI}=\lambda_2||
{\bf z}^{pred}-{\bf z}^{obs}||^2$
 \\
&$for~ i=1:N_{FLI}$&\\
 &
&  ${({\bf w^*},{\bf z^*})}=argmin_{\bf w}=\epsilon^{ML}$, then ${\bf m}^*=argmin_{\bf m}\epsilon^{FLI}$
\\   \hline
&&\\
Parallel& $(\lambda_1,\lambda_2,\lambda_3)=(>0,>0,>0)$ & ${\bf H_w}$ = Encoder-Decoder
    \\
 PIML
&$for~ i=1:N_{it}$&\\
 &
&  $\epsilon^{ML}=\lambda_1||{\bf H_w} {\bf d^{obs}}- {\bf m^{ML}}||^2;\epsilon^{FWI}=\lambda_2 ||
\overbrace{{\bf d}^{pred}}^{{\bf L m}}-{\bf d}^{obs}||^2$
    \\
&&\\
 &
&   $\epsilon=\epsilon^{ML}+ \epsilon^{FWI}+ \epsilon^{const.}\rightarrow
({\bf m}^*,{\bf w}^*)=argmin_{\bf m, w}\epsilon$
\\   \hline
&&\\
Iter. Seq.& $(\lambda_1,\lambda_2,\lambda_3)=( 0,>0,0)$& ${\bf H_w}$ = Encoder-Decoder
    \\
 PIML
& $for~ i=1:N_{it}$&\\
 &
&  ${\bf m^{ML}}={\bf H_w} {\bf d}^{obs}$;~${\bf d}^{pred} ={\bf Lm^{ML}}=\bf {L H_w d^{obs}}$
    \\
&&\\
 &
&
 $\epsilon^{FWI}=\lambda_2 ||{\bf d}^{obs}-{\bf d}^{pred}||^2\rightarrow ({\bf m}^*,{\bf w}^*)=argmin_{\bf w}\epsilon^{FWI}$
\\
\hline
&&\\
& ~$(\lambda_1,\lambda_2,\lambda_3)=( 0,>0,0)$ & ${\mathcal L} =$ PDE; ${\bf H_w}~ \& ~{\bf L_v} $ = Fully-Connected NNs
    \\
    PINN
&$for~ i=1:N_{it}$&\\
 &
&  $\epsilon^{physics}=\lambda_2 ||\mathcal{L}(\overbrace{{\bf L_v}({\bf x, m^{ML}})}^{\bf d^{pred}})-{\bf f}||^2$;~${{\bf m^{ML}}={\bf H_w} {\bf d^{pred}}}$
\\
&&
\\
&&
 $\epsilon^{FWI}=\lambda_2||{\bf d}^{pred}-{\bf d}^{obs}||^2
 \rightarrow ({\bf m}^*,{\bf v}^*,{\bf w}^*)=argmin_{{\bf v, w}}{\epsilon}^{FWI}$
\\   \hline
\end{tabular}
\end{minipage}
}
\end{table}

The PIML algorithm falls in the middle part of the spectrum in Figure~\ref{ch.ML.Inv4}, where
PIML combines physics constraints and training for a subset of the data.
 Once the ML operator ${\bf H_w}$
 is trained on a small fraction of the recorded data, then it can be efficiently applied to the rest of the data to get the optimal velocity.
 This assumes that the test data are similar to the training data.

\subsection{Benefits of PIML}
There are several potential benefits of PIML compared to standard FWI
computed by finite-difference solutions to the wave equation.

\begin{figure}[!ht]
\centering
\includegraphics[width=1\columnwidth]{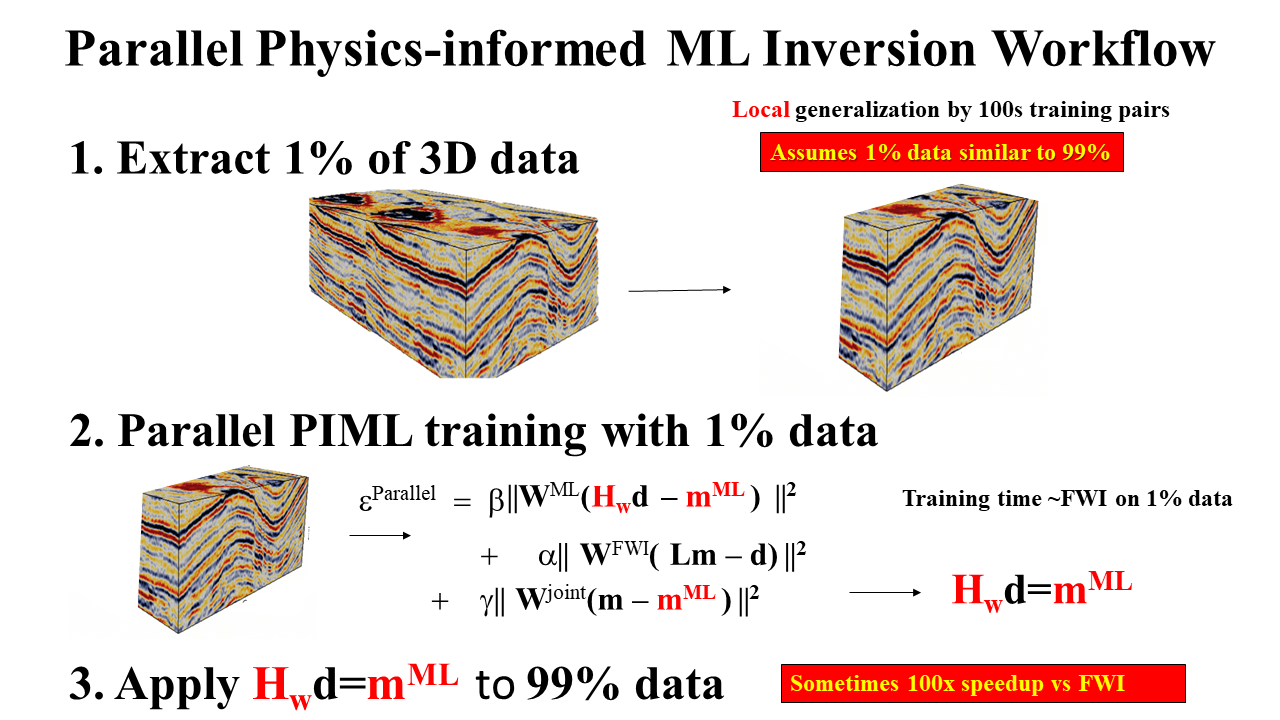}
  \caption{Workflow for the parallel PIML inversion
  of seismic data. The key assumption is that the test data are similar
  to that of the training data. This approach will likely encounter difficulties
  when the geology model has rapid velocity changes between different shot locations, such as over salt bodies in the Gulf of Mexico.}
\label{Workflow}
\end{figure}

\begin{enumerate}
\item Empirical
tests suggest that PIML algorithms constrained by the physics of wave propagation
can sometimes resist getting stuck in a local minima compared to standard FWI.

\item Compared to standard FWI, PIML can sometimes provide
much faster model estimation
of the models associated with the
recording site.
Instead of training the neural network operator ${\bf H}^{ML}$ on millions of training pairs, the training is only on a small portion, e.g. 1\%, of the recorded data as illustrated
in steps 1-2 of Figure~\ref{Workflow}.
The velocity models in the training are
computed by a traditional FWI code, with the misfit function $||{\bf W}^{FWI}({\bf L m - d})||^2$
as illustrated
in step 2 of Figure~\ref{Workflow}.
This training is typically not much more expensive than
applying FWI to this reduced data set. If the remaining 99\% of the test data
${\bf d}^{test}$ are similar to the trained data, then
the trained ${\bf H}^{ML}$ inversion operator can be used to efficiently invert
${\bf d}^{test}$
for the velocity model ${\bf m} \approx {\bf H}^{ML} {\bf d}^{test}$ everywhere
below the recording site. This is efficient because applying ${\bf H}^{ML}$
to test data can be more than an order-of-magnitude more
efficient than FWI. Training is only
performed for a small portion of the recorded
data, it is not
a global generalization for all types of data.
However, the limiting assumption is that the test data should be similar to the trained data, which
is a challenge for complex environments.

\item To overcome this limiting assumption one can
expensively pretrain a PIML architecture with data recorded
over an entire region,
and use the trained weights as
the starting weights for PIML inversion
of new data.
This is similar to the strategy of training a
large language model
(LLM) with millions of training pairs, and then using the trained weights to
produce output from new unseen queries.
These trained weights can be transfer learned
to quickly train the LLM from
more detailed information as it becomes available.
\end{enumerate}

\subsection{Previous related work}

For the geophysical inversion problems,~\cite{li2020coupled} presented a physics-informed
method for subsurface property estimation using seismic data, combining full-waveform
inversion, subsurface flow processes, and rock physics models, and using automatic
differentiation for superior accuracy and performance.~\cite{zhu2021general} introduced
ADSeismic, a seismic inversion framework that employs reverse-mode automatic differentiation
to efficiently calculate gradients.~\cite{song2021wavefield} trained a neural
network to provide flexible and continuous functional approximations of solutions without
matrix inversions, and concludes that PIML's data-constrained loss function allows it to sometimes
reconstruct a wavefield that simultaneously fits recorded data and approximately satisfies the Helmholtz
equation.~\cite{liu2023joint} proposed a differentiable physics model for large-scale joint
inverse problems for geologic carbon sequestration monitoring, which utilizes neural networks
for model variable reparameterization and forward model implementation.
They claim that this approach accurately characterizes subsurface reservoirs,
it identifies the migration of CO$_2$ plume, and it quantifies global parameters that are uncertain in the forward
models. They state: {\it ... the model can be easily deployed to high-performance
computing platforms, thereby providing a computationally efficient approach for large geophysical data}.
Another approach is by~\cite{zhang2023seismic} who incorporated both seismic data and physical law through regularization of the
loss function to estimate velocity and density fields for seismic inversion.

~\cite{karniadakis2021physics} review some methods
where physics constraints are embedded in the ML architecture.
In this case, training of the NN does not require modeling of the output by, for example, a standard finite-difference algorithm
because the trained NN finds the field values that minimize a loss function that satisfies the associated PDE~\citep{raissi2019physics}.~\cite{rotskoff2020learning} used a physics-loss function to examine transitions between two metastable states in a high-dimensional probability distribution, incorporating a variational formula for the committor function and a soft penalty on boundary conditions.~\cite{patel2022thermodynamically} developed the control volume physics-informed neural network that expands traditional finite-volume methods to deep learning environments, enabling the estimation of equations of state for shock hydrodynamics models suitable for materials like metals. PIML has also found applications in inverse and ill-posed problems in quantum chemistry~\citep{pfau2020ab}, material sciences~\citep{shukla2020physics}, and molecular
simulation~\citep{behler2007generalized}.

This paper reviews several approaches to unsupervised PIML inversion of seismic data,
where
 a traditional approach, such as a finite-difference or finite-element
method, is used to model the data.
We also  review the PINN method described by
~\cite{karniadakis2021physics}, where NN architectures are used for
both forward and inverse modeling.
The goal is to avoid the high
computational expense of computing finite-difference solutions to the 3D wave equation.
Thus far, this goal has not yet been fully realized and is one of the ongoing efforts in PINN research.

\section{Four strategies\index{Physics-informed ML} for PIML}
We now present four types of PIML algorithms: skeletal, parallel, iterative-sequential PIML,
and the physics-informed neural network (PINN).
In the first three cases, a traditional modeling method such as a finite-difference solution to
the wave equation is used to estimate the modeled data.
The fourth case of PINN, uses  NN architectures
to perform both forward modeling and inversion of the  data.
The theory for each method is described, and then numerical results are shown for
inverting geophysical data.

\subsection{Skeletal PIML inversion\index{Sequential PIML}}
\begin{figure}
\includegraphics[width=1.0\textwidth]{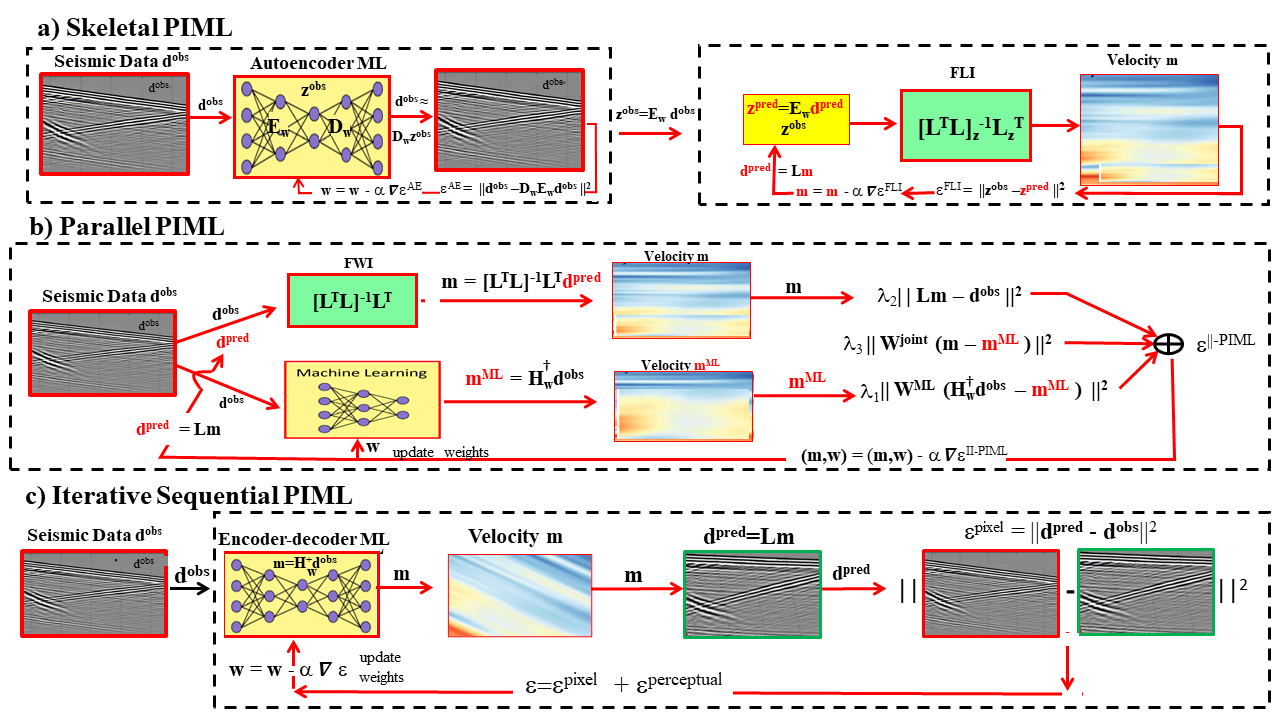}
\caption{a) Skeletal PIML strategy where the autoencoder is trained on a large
data set to give the autoencoder parameters $\bf w$ and the low-dimensional vector ${\bf z}$ that
skeletonizes the data. The full latent-vector
inversion (FLI) algorithm uses solutions to the wave equation
 to find the optimal velocity model $\bf m$
 that minimizes the sum of the squared latent-vector
 residuals $||{\bf z}^{obs} - {\bf z}^{pred}||^2$.
b) Parallel PIML strategy where the regularization term $||{\bf W}^{joint}({\bf m} - {\bf m}^{ML}) ||^2$
encourages both the ML velocity model ${\bf m}^{ML}$ and FWI model $\bf m$ to agree in parallel
with one another.
c) IS-PIML where the ML ${\bf H}_{\bf w} {\bf d}^{obs}$ and wave equation $\bf Lm=d$ modeling
operators are sequentially executed because, unlike the parallel PIML strategy, the ML model is
equated to the actual velocity model for forward modeling.}
\label{ch.ML.Inv3}
\end{figure}

To reduce both the dimension and complexity of the input to a wave-equation inversion algorithm,~\cite{chen2020seismicb} and~\cite{chen2020seismica} introduced a hybrid ML and physics-based inversion strategy denoted as {Newtonian Machine Learning} (NML).
Figure~\ref{ch.ML.Inv3}a illustrates their strategy, where an autoencoder (AE)
is trained to condense
important features of input seismic traces ${\bf d}_i~for ~i\in[1,2,...,M] $ into
small-dimensional latent vectors ${\bf z}_i~for ~i\in[1,2,...,M]$.
The latent vectors are then inverted for the velocity model
by finite-difference solutions to the wave equation, as summarized by
the top row of Table 1.

The weights for the AE are defined by
\begin{eqnarray}
{\bf w}^*&=& argmin_{{\bf w}} \sum_{i}^M ||~\overbrace{{\bf H}_{\bf w} {\bf d}_i^{obs}}^{predicted~data~
 ~{\bf d}_i^{pred}}-{\bf d}_i^{obs}~||^2,
\label{PIML.AE.eq1}
\end{eqnarray}
where ${\bf H}_{\bf w}=\bf D_{\bf w} {\bf E}_{\bf w}$ represents the forward
 modeling operator of the AE,
$\bf E_{\bf w}$ and $\bf D_{\bf w}$ are the respective encoder and decoder operators,
and ${\bf w}^*$ represents the trained weights of the AE. After training, the $i^{th}$ observed latent vector
${\bf z}_i$ associated with ${\bf d}_i$ is
obtained by
\begin{eqnarray}
{\bf z}_i&=&{\bf E}_{{\bf w}^*} {\bf d}_i~for~i=[1,2,...,M].
\label{PIML.AE.eq2}
\end{eqnarray}
We then use forward modeling of the wave equation to produce $M$ predicted traces ${\bf d}_i^{pred}$ for a
trial-velocity model, which are then used to compute the $M$ predicted latent vectors
of low dimension:
\begin{eqnarray}
{\bf z}_i^{pred}&=&{\bf E}_{{\bf w}^*} {\bf d}_i^{pred}~for~i=[1,2,...,M].
\label{PIML.AE.eq2a}
\end{eqnarray}
The trial-velocity model $\bf m$ is adjusted until all the predicted latent vectors ${\bf z}_i^{pred}$
match the observed ones ${\bf z}_i$ in a least squares sense.
That is,
\begin{eqnarray}
{\bf m}^* &= argmin_{\bf m} ~\sum_{i=1}^M \overbrace{||\underbrace{{{\bf z}_i^{pred}} - {\bf z}_i}||^2 ~+~regularization}^{\epsilon^{FLI}},
\label{PIML.AE.eq4}
\end{eqnarray}
where ${\bf m}^*$ is the optimal velocity model that minimizes the regularized
sum of the  squared latent-space residuals $\epsilon^{FLI}$. Here, we define
 $\epsilon^{FLI}$ as the objective function
for full latent-vector inversion (FLI).
A gradient-descent formula with step-length $\alpha$ is used to iteratively find ${\bf m}^*$ s.t.
\begin{eqnarray}
{\bf m} &\defeq & {\bf m} - \alpha \nabla_{\bf m} \epsilon^{FLI},
\label{PIML.AE.eq4a}
\end{eqnarray}
where $\nabla_{\bf m} \epsilon^{FLI}$ is computed using finite-difference solutions to the
wave equation~\citep{chen2020seismicb,yu2021skeletonized}.
The backpropagation of residuals can be performed by a code
that honors the explicit equations of a gradient derived by~\cite{chen2020seismicb},
or backpropagation of residuals can be computed by automatic differentiation.

In summary, skeletal PIML consists of the two sequential steps described in the top row of Table 1: training of the AE on the observed traces
to get the latent-space vectors ${\bf z}_i$,
then use wave-equation solutions to find the optimal velocity model $\bf m$
 that minimizes
the sum of the squared latent-vector residuals. The first step is unsupervised AE training and the
second step uses the physics of wave propagation to invert for the velocity model.
This is equivalent to setting $\lambda_1=1$ and $\lambda_2=0$ in
equation~\ref{ch.InvIntro.eq1aa} for all of
the iterations until convergence. Then set $\lambda_1=0$ and $\lambda_2=1$ in
equation~\ref{ch.InvIntro.eq1aa}, and use the latent variables as the observed data for the FWI objective function. An iterative gradient descent
method is then used to find the velocity model that minimizes
the z-domain misfit function in equation~\ref{PIML.AE.eq4a}.

\subsubsection{Numerical example}
As an example,~\cite{yu2021skeletonized} tested skeletal PIML on refraction data recorded next
 to the Gulf of Aqaba on the Saudi Arabian
 peninsula.
 A 12-lb hammer striking a plate on the ground is used as
 the seismic source, and each source excitation was recorded by 120 vertical-component
 receivers placed along a line. The source and receiver spacings are 2.5 m, where
 each source is excited next to a receiver position.

\begin{figure}[!ht]
\centering
\includegraphics[width=1\columnwidth]{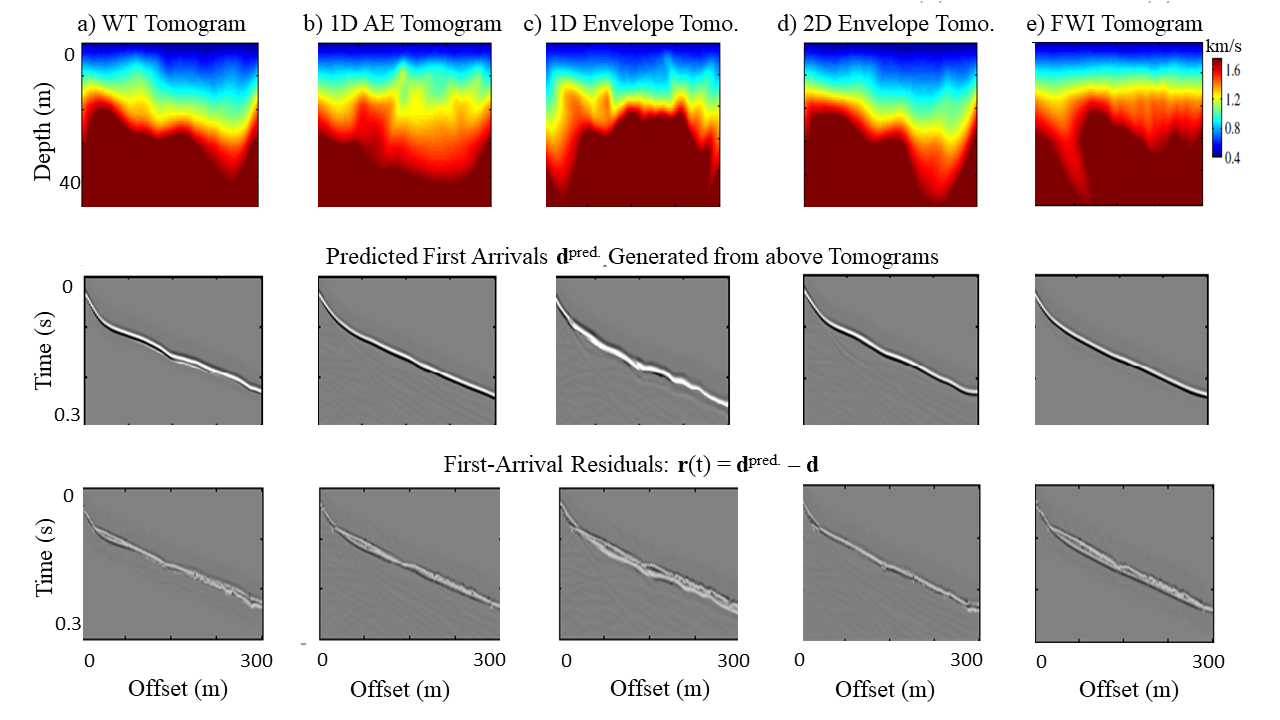}
  \caption{Top row are tomograms inverted by a) the 50-Hz wave equation
  traveltime (WT) inversion method~\citep{luo1991waveb},
  b) AE inversion with a 1D latent
  space and a 50-Hz bandpass filter
  applied to the data, c) envelope inversion,
  d) AE inversion with a 2D latent space and e) FWI. Figure from~\cite{yu2021skeletonized}.
  Below each tomogram are the corresponding first arrivals
  ${\bf d}^{pred.} $ generated from the tomogram's velocity model and
  the associated residual traces ${\bf r}={\bf d}^{pred.}-{\bf d}$ where $\bf d$ represents
  the recorded trace.
  Figure adapted from~\cite{yu2021skeletonized}.}
\label{AE.Aqaba6}
\end{figure}

~\cite{yu2021skeletonized} used
 an auotoencoder to reduce the dimension of the input seismic traces, and
 then inverted a small-dimension latent space vector for the velocity model.
 The inversion strategy was that of skeletonized inversion which used  finite-difference
 solutions to the wave equation.
 For comparison, a
variety of different methods are used to invert the first-arrival data for
the velocity tomograms in Figure~\ref{AE.Aqaba6}.
The wave-equation traveltime inversion (WT) method is used to
compute Figure~\ref{AE.Aqaba6}a, where the input windowed data
are bandpass filtered peaked at 50 Hz. The starting model is
a linear gradient velocity model~\citep{yu2021skeletonized}, and the
same data are inverted by the NML method to give
the tomogram in Figure~\ref{AE.Aqaba6}b.
The envelope
inversion results are shown in  Figure~\ref{AE.Aqaba6}c
and the FWI\index{Full waveform inversion (FWI)} tomogram is displayed
in Figure~\ref{AE.Aqaba6}e.
These results were obtained after 20 iterations.
The first four inverted
tomograms, especially the first and the fourth tomograms
in Figure~\ref{AE.Aqaba6}a and Figure~\ref{AE.Aqaba6}d,
suggest the possible location of the Aqaba
fault scarp~\citep{hanafy2014imaging} observed on the surface.

The reconstructed early arrivals ${\bf d}^{pred.}$ from each of the tomograms in
Figure~\ref{AE.Aqaba6} are displayed in the second row of Figure~\ref{AE.Aqaba6}
and are computed by a finite-difference
modeling method. Here, the predicted arrivals
are subtracted from the actual arrivals to get the residual
CSG\index{Common shot-point gather (CSG)}s in the third row of Figure~\ref{AE.Aqaba6}.
It is clear that the 2D NML residual in Figure~\ref{AE.Aqaba6} has the smallest
residual compared to the other methods.

\begin{figure*}[!ht]
\centering
\includegraphics[width=1\columnwidth]{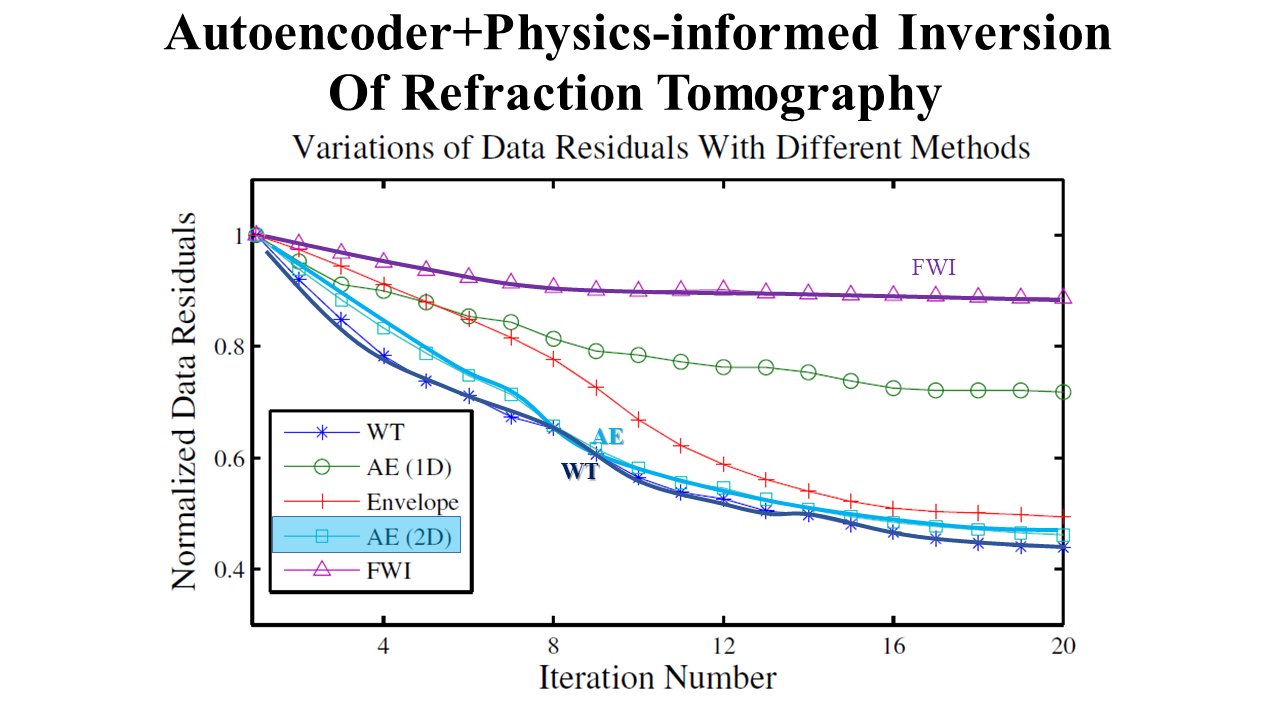}
  \caption{Iteration vs residual for the five
algorithms presented in Figure~\ref{AE.Aqaba6}. Figures adapted from~\cite{yu2021skeletonized}.}
\label{AE.Refract}
\end{figure*}

The data misfit curves with the five
algorithms are presented in Figure~\ref{AE.Refract}.
These curves show that the WT and 2D AE
methods converge the quickest to a small data residual.

\subsection{Parallel PIML inversion\index{Parallel ML+Physics inversion}}
Instead of the skeletal use of ML training followed by
a physics-based inversion algorithm,
equation~\ref{PIML.eq120} and
Figure~\ref{ch.ML.Inv3}b illustrates
the parallel minimization~\citep{colombo2021framework,colombo2021physics}
of the
ML and physics-based objective functions
in equation~\ref{PIML.eq1}.
Here, the matrices $({\bf W}^{FWI}, {\bf W}^{ML})$
are used to downweight unreliable parts of the data or model parameters, and the joint misfit term
is used to encourage the ML and physics-driven
models to be identical. The positive scalar variables
$(\lambda_1, \lambda_2, \lambda_3)$ in the second row of Table 1 moderate the relative
importance of the three misfit functions.

A gradient-descent strategy
with alternating updates of the ML parameters ${\bf m}^{ML}$ and velocity model
${\bf m}$ can be used
to find the optimal velocity model ${\bf m}^*$.
First, the model ${\bf m}$ is fixed and the parameters for the NN
are computed to minimize the objective function $\epsilon^{||-PIML}$ in equation~\ref{PIML.eq120}.
The parameters for the ML algorithm can be
initially trained offline on
a training set consisting of seismic data and their associated
velocity models.
This is therefore classified as a supervised learning method.
 Then the updated ML parameters are fixed and the model
 ${\bf m}$ is found by gradient descent\index{Gradient descent (GD)} that minimizes
$\epsilon^{||-PIML}$. The term $||{\bf W}^{joint}({\bf m}-{\bf m}^{ML})||^2$
 encourages agreement between the  ${\bf m}$
 and ML ${\bf m}^{ML}$ models. This alternating
 use of the different objective functions
 is repeated until acceptable convergence.
 Examples of
this procedure for geophysical inversion are presented in~\cite{colombo2021framework,colombo2021physics}. Other
physical information, such as seismic faces~\citep{li2021deep,zhang2022regularized},
initial models~\citep{zhang2021robust}, spatial-temporal information~\citep{yang2022making}, and
migration images~\citep{zhang2020adjoint}, can be also used as the regularization term in the
objective function to improve the inversion results.~\cite{colombo2023machine} also developed a physics-adaptive ML inversion scheme
that shows good generalization properties for field applications.
Their deterministic inversion scheme is regularized by a penalty term
characterized by the difference between the inverted models and ML-predicted models.
The resulting models and computed responses are used for augmenting the training data.

\begin{figure*}[!ht]
\centering
\includegraphics[width=1\columnwidth]{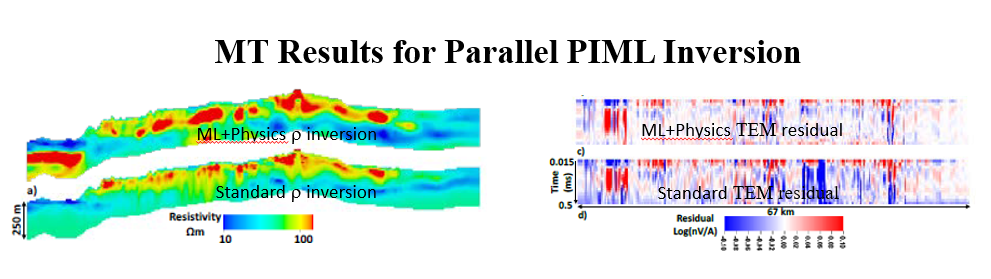}
  \caption{Left column of images compare the results for inverting the resistivity model
  $\rho({\bf x})$ from TEM data using the parallel PIML method and the standard
  pure-physics inversion. The residuals of the TEM data are plotted on the right. Figures adapted from~\cite{colombo2021physics}.}
\label{AE.Aqaba88}
\end{figure*}

~\cite{colombo2021physics} used the parallel PIML method to invert TEM data. The reconstructed resistivity
model is shown in the left column of Figure~\ref{AE.Aqaba88}, where the resistivity model
obtained by a standard pure-physics algorithm is shown just below it.
Both models look similar. However, if the data generated by these models are synthetically
computed then the residuals for the parallel PIML model are displayed in the right column of
Figure~\ref{AE.Aqaba88} along with  those from the standard pure-physics model. It is clear
that the parallel PIML results
are more accurate than that from standard inversion.

\subsection{Iterative-sequential PIML inversion\index{Iterative-sequental PIML inversion}}
The parallel PIML ican sometimes require offline training
of the ML weights using data and model pairs
generated by, for example, synthetic
simulations. These simulations should be  for models
that resemble the actual model that is to be reconstructed from the recorded data.
This can be tedious and
time consuming  for large training sets. To avoid this problem,~\cite{jin2021unsupervised}
developed the unsupervised
physics-informed ML inversion method. In this case, the velocity models
are obtained by an encoder-decoder network as shown in  Figure~\ref{ch.ML.Inv3}c.
Here,
\begin{eqnarray}
{\bf w}^*&=& argmin_{{\bf w}} [ \overbrace{\sum_{i}^M ||~ {\bf d}_i^{obs}-{\bf d}_i^{pred}~||^2 }^{\epsilon^{pixel}}+\epsilon^{perceptual}~],
\label{PIML.AE.eq7}
\end{eqnarray}
where ${\bf d}_i^{obs}$ (${\bf d}_i^{pred}$) represents the $i^{th} $ observed (predicted) trace, ${\bf  H}_{\bf w}= {\bf D}_{\bf w}  {\bf E}_{\bf w}$
represents the forward modeling operator of the encoder $\bf E_{\bf w}$ and decoder $\bf D_{\bf w}$ operations,
and ${\bf w}^*$ represents the trained weights of the encoder-decoder architecture.
The perceptual loss $\epsilon^{perceptual}$ objective function is
\begin{eqnarray}
\epsilon^{perceptual}&=&\sum_i (F({\bf d}_i^{obs})-F({\bf d}_i^{pred}))^2,
\end{eqnarray}
which is used to form the complete objective function
\begin{eqnarray}
\epsilon&=&\epsilon^{pixel}+\epsilon^{perceptual}.
\end{eqnarray}
Here,
$F({\bf d}_i)={\mathbf z}_i$ is the latent-vector
output ${\mathbf z}_i$ of an encoder network trained on a set of data.
These data are those of images and the encoder associated
with the training is that of a VGG network~\citep{johnson2016perceptual}.

Unlike the autoencoder where the input and output are the recorded traces,
${\bf  H}_{\bf w} {\bf d}^{obs}=\bf m$ produces the velocity model $\bf m$ from
the recorded data ${\bf d}^{obs}$.
This velocity model then produces the predicted shot gather ${\mathbf d}^{pred}$
shown in Figure~\ref{ch.ML.Inv3}c, which is obviously in disagreement
with the input shot gather ${\bf d}^{obs}$ at the far left.
This is because the velocity model generated by the
autoencoder is incorrect
because its weights $\bf w$ at early iterations do not reproduce a velocity model
that predicts the actual data.

A gradient-descent formula
\begin{eqnarray}
{\bf w} = {\bf w} - \alpha \nabla \epsilon,
\end{eqnarray}
is then used to adjust the weights of the encoder-decoder so that a more accurate
velocity model ${\bf m}$ is generated in Figure~\ref{ch.ML.Inv3}c.
This computed velocity model
is then forward modeled ${\bf d}^{pred}={\bf L m}$ to get the predicted shot gathers ${\bf d}^{pred}$
by a numerical solution to the wave equation.
Repeating these sequential series of steps
until $\epsilon$ is minimized
defines the IS-PIML method.
Some  benefits of this method are that it is unsupervised
and   once it is properly trained,
 new input data can be quickly inverted by the encoder-decoder to get the velocity model.
 The third row in Table 1 summarizes these procedures.

In practice, the gradient for small batches in each epoch  leads to a
lower computational cost than pure physics-based inversion. However,
it calculates the data loss through physics-based forward modeling, which leads to a
higher cost than the supervised methods~\citep{deng2021openfwi}.

\subsubsection{Numerical example}
\begin{figure*}[!ht]
\centering
\includegraphics[width=1\columnwidth]{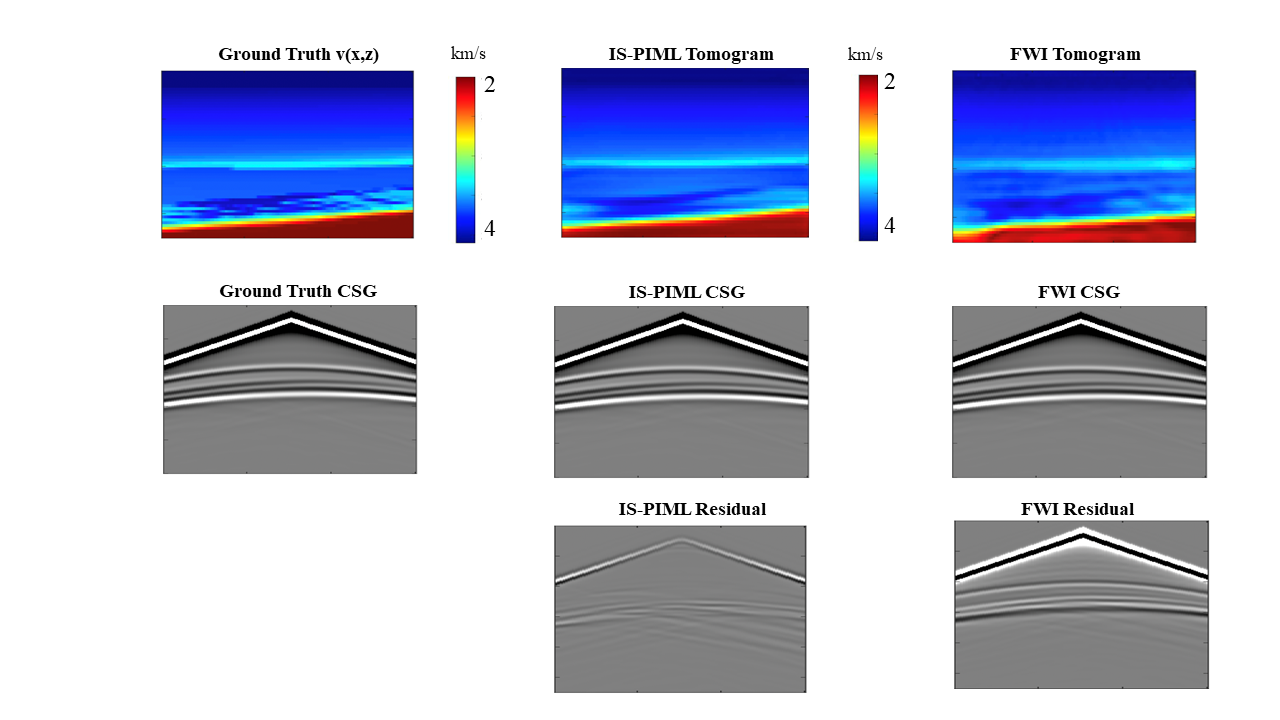}
  \caption{Comparisons of ground truth velocity model and tomograms
  (top row), predicted CSGs (middle row) and residuals
  (the last row)
  computed by the IS-PIML and FWI algorithms.}
\label{UPFWIa}
\end{figure*}

A synthetic  data set is used to test the effectiveness of
the IS-PIML method. A finite-difference method is used
 to generate synthetic data from the true velocity model
 in the left column of Figure~\ref{UPFWIa}.
 These data are inverted to give the IS-PIML and FWI tomograms in the middle
 and right columns, respectively.
A finite-difference method is then used to generate the predicted
CSGs using the velocity models from the tomograms, and these predictions are subtracted from the true traces
to give the residual traces at the bottom row of
Figure~\ref{UPFWIa}~\citep{feng2022exploring}.
The IS-PIML residuals have noticeably less residual energy compared to the FWI result.
In terms of computational cost, the   IS-PIML algorithm costs several times more than the FWI for this example.

\subsubsection{Multiphysics PIML}

The PIML algorithms can be used to invert for multiple types of
data, each governed by a different type of physics. This can be done
by creating a joint objective function by regularizing and summing together
weighted objective functions associated with each type of data.
Typically, the regularization function enforces a soft constraint
that equates one type of model with the other one.

\begin{figure*}[!ht]
\centering
\includegraphics[width=1\columnwidth]{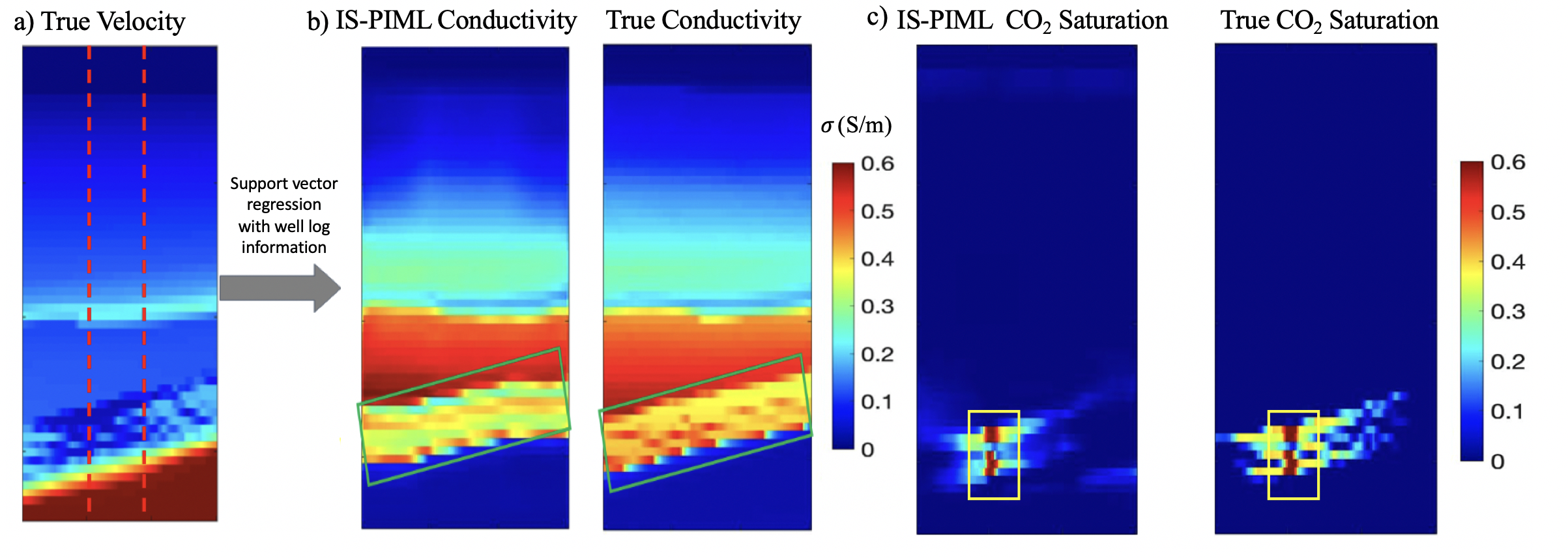}
  \caption{a) Well logs denoted by red dashed lines,
  b) IS-PIML predicted and true conductivity.  c) IS-PIML predicted and true CO$_2$ saturation.}
\label{UPFWIb}
\end{figure*}

For example, consider the seismic velocity model in Figure~\ref{UPFWIb}a
where the dashed red lines correspond to well locations. Conductivity
logs $\sigma(z)$ and $CO_2$ saturation logs $s(z)$ are recorded and are correlated to the velocity tomogram at each depth
using a support regression algorithm~\citep{feng2022exploring}.
This gives the mapping between the seismic velocity values
and the conductivity  $F(v(x_{well} ,z))=\sigma(z)$ and $CO_2$ saturation
$G(v(x_{well} ,v(z))=s(z)$ values, where $v(x_{well} ,z)$
is the vertical velocity profile of the tomogram
 at the well offset $x_{well}$. These functions $F(v)$ and $G(v)$
are used to map the 2D tomogram velocity $v(x,z)$ to the pseudo-conductivity $\sigma(x,z)$ and the pseudo-CO$_2$ saturation.
With the pseudo labels, the iterative sequential-PIML methodology gives predicted conductivity and CO$_2$ saturation in Figure~\ref{UPFWIb}band~\ref{UPFWIb}c. Both images compare well to the true conductivity and CO$_2$ saturation
images.

\begin{figure*}[!ht]
\centering
\includegraphics[width=1\columnwidth]{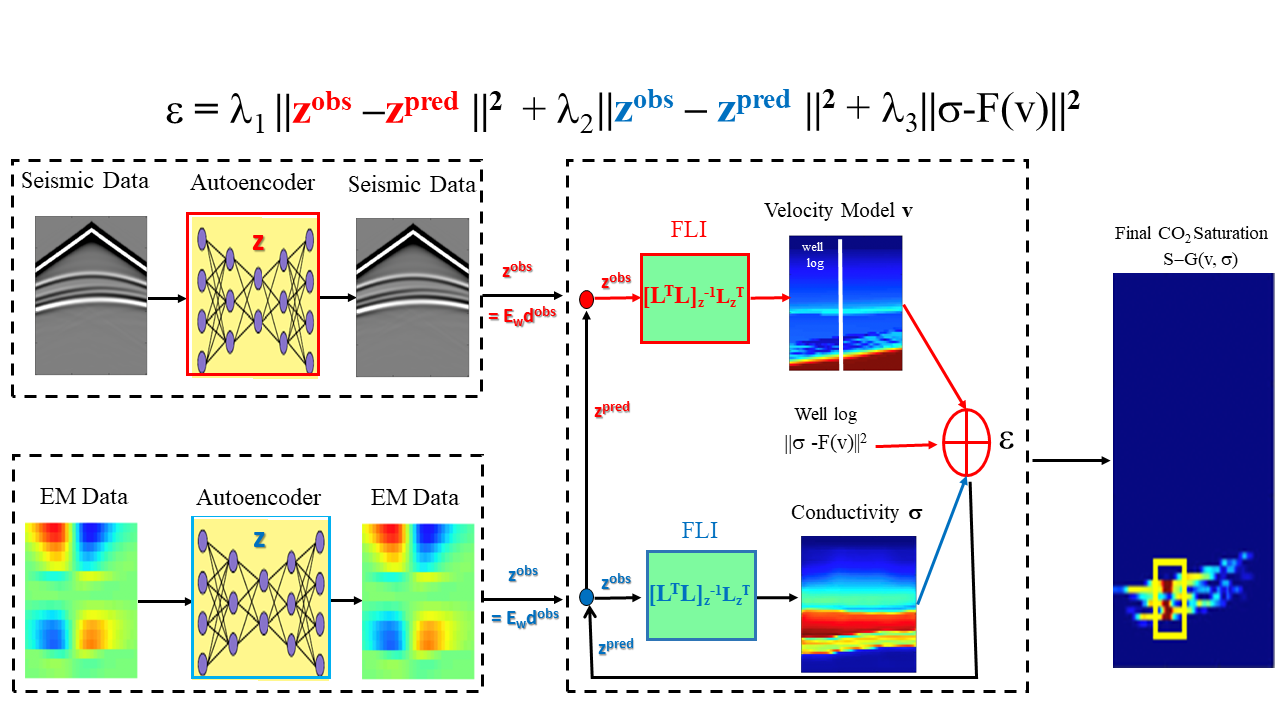}
  \caption{Hypothetical schematic for multiphysics sequential PIML where electromagnetic (EM) and seismic
  are skeletalized to latent vectors $\bf z$, and the output consists of conductivity
  and velocity images inverted by FLI. These images are constrained by the mapping $\sigma=F(v)$
  between conductivity and velocity obtained
  from well-log information. The saturation mapping from conductivity and
  velocity values is also obtained from well logs to get $S=F(c,\sigma)$, which is used to
  get the CO2 saturation image on the right.}
\label{AE.Multi}
\end{figure*}

The sequential PIML can also be used to combine models from different
types of data and well-log information to obtain models constrained
by well-log information.
Figure~\ref{AE.Multi} provides an example of this for electromagnetic (EM) and seismic data
to estimate the pseudo-saturation image.
 The joint
  objective function at the top is a combination
  of FLI misfit functions
  and the well-log constraint.

\subsection{Physics-informed neural network}

The previous sections showed different ways
of combining an objective function
computed by a NN and one
computed using a finite difference solution
to the wave equation.
Now we introduce a {\it physics-informed neural network} (PINN) that only uses
NN architectures to compute both the forward modeling solution
 to the wave equation as well as its inverse solution~\citep{raissi2019physics,karniadakis2021physics}. The hypothetical advantage
is that PINN can hopefully avoid the large computational expense of finite-difference
modeling of 3D data, with the possible penalty of reduced accuracy. This goal has not yet been fully met,
and it is too early to judge the practical benefits of PINN compared to FWI in inverting seismic data~\citep{youtube}.
The PINN schematic for inverting data can be represented by
the IS-PIML schematic in Figure~\ref{ch.ML.Inv3}c, except the
forward modeling operator $\bf L$ is now computed by a trained NN
rather than a FD solution to the wave equation.

Appendix I  describes how PINN solves the forward problem
by an NN, and then explains the strategy for inverting for model parameters from recorded data.
Examples for modeling and inverting synthetic data by
 a NN are described in the next section, and the PINN algorithm is summarized in the last row of Table 1.

\subsubsection{Examples of inverting seismic data by PINN}

\begin{figure*}[!ht]
\centering
\includegraphics[width=1\columnwidth]{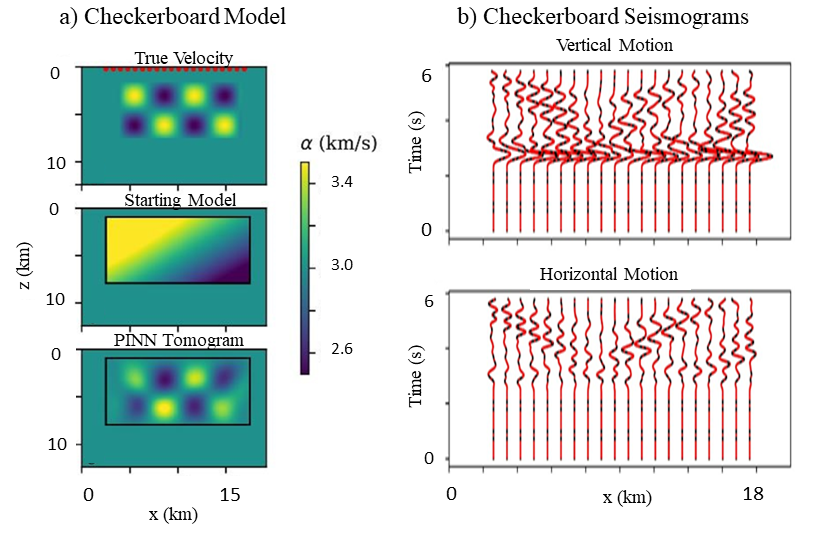}
\caption{(a) True velocity, starting velocity and inverted velocity models (aka tomogram),
and (b) corresponding predicted and true seismograms recorded along the top
of the model. The blue lines are the true seismograms computed by a type of a finite-element method
and the red dashed lines are computed by the PINN.}
\label{Fig1:PINN.300}
\end{figure*}

The checkerboard velocity model
in Figure~\ref{Fig1:PINN.300} is used to demonstrate
PINN inversion of acoustic data for the velocity model. Here, the true velocity model is shown at the top
of (a), the starting model is in the middle of (a)
and the bottom image is the tomogram inverted by PINN~\citep{rasht2022physics}.
The tomogram and the actual model largely agree with one another.

In this example,
there are 20 evenly-spaced seismometers (red squares) along the top of the model
and there are nine-point sources evenly distributed
at a depth of 10 km, each having a Gaussian wavelet with a dominant frequency of 20 Hz
and the recording time is 5 s.
Therefore, this example resembles that for a crosswell experiment with horizontal source and receiver wells.
The free surface boundary condition for this simulation was not enforced.

The NN consists of 10 layers, each with 100 neurons per layer.
There are 11600 total samples in all of the recorded seismograms,
and the training data for the PDE and early-arrival snapshots consist of 60,000 and 3,600 points respectively.
There are  two early time snapshots (at $t=0$ and $t=0.05 ~s$) in the training data, and
they are used
as enforced data constraints that get information
about the background velocity.
There is a typo in the~\cite{rasht2022physics} paper, and the author (Huber, pers. comm.) states that the
data residual in the loss function is actuated and the sources are at a depth of 10 km.

The
actual and predicted seismograms in Figure~\ref{Fig1:PINN.300}b) mostly agree with one another.
The physics loss term is computed on a set of randomly selected training points $(x_1, x_2,t)$, and
the loss term associated with two early snapshots is weighted by a factor that is 10 times more than the
physics loss term.
~\cite{rasht2022physics} state that
the current state-of-the-art PINNs provide good results for the forward model, even though spectral element or finite-difference methods are more efficient and accurate.

Some recent improvements that improve the performance of PINN  include the following.
\begin{enumerate}
\item {\bf Adaptive activation function.} The $\tanh(x)$ activation sometimes fails to
capture high-frequency solution modes,
so it is replaced by an adaptive activation function such as
$\tanh(\alpha x)$ or $max(0,\alpha x)$, where $\alpha$ is a learnable parameter for each layer. This can help capture finer
details in the predicted data and the model. Empirical results suggest that the siren function $\sin(x)$
can sometimes be even more effective in capturing finer details.
Employing periodic activation functions also increased the effectiveness in
predicting pressure and solute concentration fields in
heterogeneous porous media~\citep{faroughi2022physics}.

\item {\bf Dynamic weighting of the loss functions.}
~\cite{wang2021understanding} observed that the NN gradient for the gradient tends to favor satisfying the physics equations at the expense
of largely ignoring the data and boundary constraints. Thus, the loss functions should be weighted to better balance the data terms for unbalanced constraints.
For example, the scalar weight $\lambda_{boundary}$
for the boundary loss term $\lambda_{boundary} \epsilon^{BC}$ should be much greater than that for the physics term $\lambda_{physics} \epsilon^{physics}$ if it is weighted as
\begin{eqnarray}
\lambda_{boundary} &=& \frac{|\nabla_{\theta} \epsilon^{physics}|}{|\nabla_{\theta} \epsilon^{BC}|},
\end{eqnarray}
because $|\nabla_{\theta} \epsilon^{BC}|$ is often much smaller than $|\nabla_{\theta} \epsilon^{physics}|$ for most inputs into an NN layer; here, $\theta$ represents the NN parameters.
Another approach is to identify which types of training data are harder to train and force the NN to focus on those data before training the next step~\cite{li2022dynamic}.

\item Random distributions of points in time and space are more effective
than a uniform gridding. The density of points should be increased where the physics loss is high~\citep{yu2021skeletonized}.

\item A quasi-Newton L-BFGS is more effective than an SGD approach.

\item
\cite{huang2022pinnup} claim that, compared to the commonly used PINN with random
initialization, their frequency upscaling approach
exhibits notable superiority
in terms of convergence and accuracy with a two-hidden-layer
model.

\item
\cite{sun2023implicit} used the PIML approach to
invert synthetic seismic data. Instead of using
a ML method for their forward modeling, they used a standard finite-difference solver so that their scheme can be described by the IS-PIML schematic in Figure~\ref{ch.ML.Inv3}c.
Their empirical results showed that the PIML inversions
were largely resistant to getting stuck in a local minimum compared to
the standard FWI approach for poor starting models.
See Figure~\ref{Fig1:PINN.305}.
They also found that replacing the ReLu functions with sine-based activation functions speeded up convergence by more than an order of magnitude. However, the reconstructed models from PIML appeared to
be less accurate compared to FWI tomograms when significant
amounts of random noise were added to the data
(see Figure~\ref{Fig1:PINN.315}).
They claimed that their approach ''exhibits a strong capacity for
generalization, and is likely well-suited for multi-scale joint geophysical inversion.''.

\begin{figure*}[!ht]
\centering
\includegraphics[width=1\columnwidth]{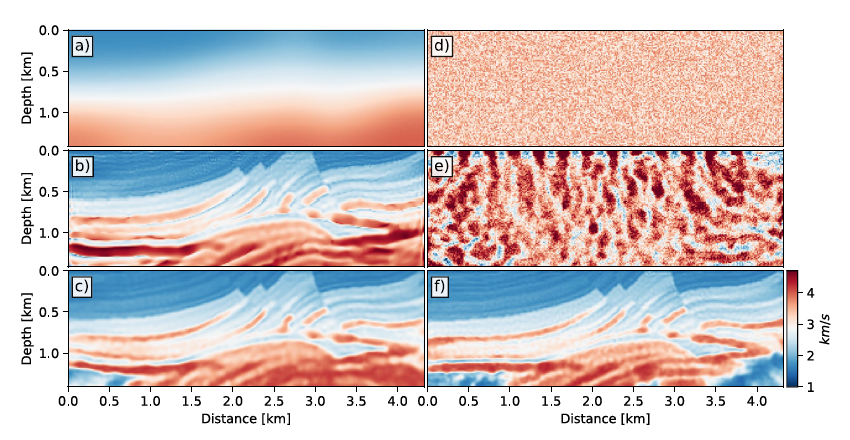}
\caption{The first row of images depicts the a) smooth and d) random
initial-velocity models.
From left to right,
the second row depicts the tomograms obtained by  FWI
for the smooth and random starting models.
The third row is the same except these tomograms are
computed by the PIML inversion in~\cite{sun2023implicit}
These images from~\cite{sun2023implicit} suggest that
PIML inversion is somewhat resilient to poor starting models.}
\label{Fig1:PINN.305}
\end{figure*}

\begin{figure*}[!ht]
\centering
\includegraphics[width=1\columnwidth]{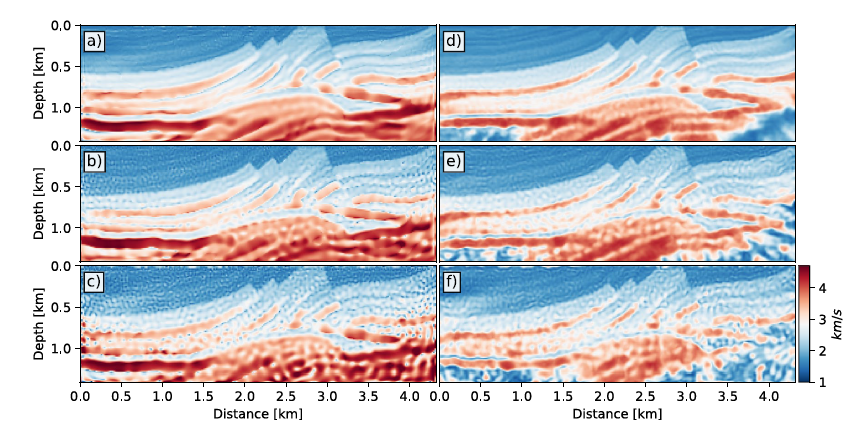}
\caption{The first row of images depict the tomograms
inverted by a) FWI and d) PINN from the noise-free data
and a good starting model.
From left to right,
the second row depicts the tomograms computed by the b) FWI and e) PINN methods where the input data
have additive random noise with $\sigma=2\sigma_0$.
The third row is the same except the input data are twice as noisy
 with $\sigma=4 \sigma_0$.
These images from~\cite{sun2023implicit} suggest that
PIML inversion is somewhat more sensitive to noisy data than FWI.}
\label{Fig1:PINN.315}
\end{figure*}

\end{enumerate}
\section{Summary}

We overviewed  four types of PIML strategies
that combine an ML algorithm with a physics-based inversion strategy.
The first three require solutions to the wave equation using a standard
solver such as the finite-difference or  finite-element methods. The last one, denoted as PINN,
only requires a neural network to find the solutions to the governing equations and the model parameters.
\begin{enumerate}

\item Skeletal PIML is an unsupervised method that
reduces the dimension of the data to the latent-space
variables, and then a physics-based inversion inverts them
to estimate the optimal model. Dimensional
reduction of the data often reduces the
chances of getting stuck in a local minimum, but
at the possible cost of losing some resolution.  There is no need for supervision or
for estimating
the regularization values  because
of the sequential one-and-done nature of each algorithm.
The unsupervised autoencoder of~\cite{ladjal2019pca} is a non-linear generalization of the linear
principal component
transform (PCT), where the singular vectors with the largest singular values
are similar to the latent vectors from an autoencoder. Their PCT autoencoder
projects data
to and from the lower-dimensional subspace
so that
the latent-space components (axes) are ordered in terms of decreasing
importance and each component of a code is statistically independent from the other components.

\item The parallel PIML
strategy is to  train the ML algorithm
and the physics-based algorithms in parallel reconstruct both
${\bf m}^{ML}$ and $\bf m$, and use the penalty
term $||{\bf m}-{\bf m}^{ML}||^2$ to bring the models
into agreement with one another.
The starting weights for the ML architecture can be trained with labeled data,
and then retrained with the parallel PIML procedure for new data.
Once the ML algorithm is sufficiently trained then
using ${\bf m}^{new} = {\bf H_w} {\bf d}^{new}$ to get
the new model  ${\bf m}^{new}$ from new data ${\bf d}^{new}$ is
typically much less costly than
that from a physics-based inversion algorithm. The assumption is that ${\bf m}^{ML}$
is close enough to the global minima so that its neighborhood
can be explored by
the physics-based gradient algorithm to quickly find the global minimum.
Determining the optimal values of the regularization parameters is an active area of research.

\item Iterative-sequential PIML  is an unsupervised method
where  the output of the ML model ${\bf m}^{ML}$
model is used to generate the input data ${\bf d}^{pred}$. This $k^{th} $ iterate data
is used to generate a data misfit function $||{\bf d}-{\bf d}^{pred}||^2$, which is then minimized by updating the ML weights
by a gradient-descent method. Once the IS-PIML is trained, then the
trained ML portion of the algorithm can be used to
efficiently predict models from new data.

Empirical
tests suggest that constraining
the ML algorithm to honor the physics of the data
can sometimes yield more efficient and accurate reconstructions of earth models
compared to pure ML-based inversion.
Another benefit is that the PIML training time for the ML algorithm
can sometimes require less computational effort compared to standard ML
that only uses data for training.
This trained ML architecture
can then be used to
economically invert for models from new data ${\bf d}$ compared to standard FWI.
Such data can be recorded over geology similar to that of the training data or include many
more unused seismic lines from the same survey.
Another benefit is that some PIML algorithms do not require labeling of the training data, which
can be a big advantage for large data sets.
Improving the robustness, convergence speed, and accuracy of these hybrid PIML strategies is an active
area of research.

\item The PINN strategy does not require a standard finite-element or finite-difference solver to
compute solutions to the wave equation, or its inverse. It only requires
a neural network, where the optimal loss function satisfies the PDE, boundary+initial conditions, and recorded data. Results so far suggest that PINNs can accurately find solutions to the wave equation for smooth velocity models. However, many more epochs are needed to find wavefield solutions for complex velocity models.
Thus far, PINNs are not more efficient than standard solvers for general velocity models. However, if the solutions to a large number of
velocity models are needed, then solutions by a trained PINN are claimed to be more efficient than those from a standard solver.
This assumes that the models are similar to one another.
One strategy is to use the trained
weights from a modest number of models to
be the initial weights for training the NN for other models.

It is too early in its development to state that inversion of the velocity model by PINNs is more effective and efficient
than standard FWI. Nobody has come close to demonstrating this claim.  There are many problems to overcome,
including the low accuracy and slow convergence with complex velocity models, optimal selection of the network size, the weighting of the misfit functions, choice of network architecture, data size, step length, type of gradient descent method, and many other issues.
\clearpage
\end{enumerate}

If the PIML approach is ever to replace standard FWI then it must
be an order-of-magnitude faster,
be just as accurate, and be just as robust with noisy and incomplete data as FWI. Thus far, this
goal has not yet been achieved.
However, if the NN modeling and inverse operators are expensively
pretrained to be generalized for data recorded over a large geographic region, then the trained weights can be transfer learned to expedite
the PIML inversion of new data collected in the trained region. For example,
a general PIML can be pretrained with
3D data over many data examples recorded in the Ghawar
region of Saudi Arabia. These weights can then be used for
PIML inversion of time-lapse data or new data
recorded in that region. This assumes that
the NN operations are much more efficient than that
of standard FWI, which is almost always the case.

This goal of pretraining a PIML architecture with data recorded
over an entire region
is similar to the strategy of training a
large language model
(LLM), and then using the trained weights to
produce output from new unseen queries.
These weights can be transfer learned
to quickly train the LLM from
more detailed information as it becomes available.

\section{Appendix I: Forward and Inverse Modeling by PINN}

This appendix first explains forward modeling by PINN. The next section describes both the forward modeling and inversion methods using PINN.

\subsection{Forward modeling by PINN}\index{Physics-informed neural networks (PINNs)}

To avoid the finite-difference modeling of the velocity model,~\cite{raissi2019physics} and~\cite{karniadakis2021physics}
introduced the PINN approach that uses the partial differential equations of the governing
equations as a constraint to the NN system. If $\mathcal{L} u=f$ represents the PDE applied
to the scalar field $u$ and $f$ is the localized source function, then the loss function is defined as the sum of the squared residuals
\begin{eqnarray}
\epsilon^{PINN} &=& \sum_{i \in {\mathcal D}} (\mathcal{L} u_i -f_i)^2 + {regularization},
\label{Inv.FWI.ML.eq2}
\end{eqnarray}
where the summation is over the points in the entire computational domain $\mathcal D$
and the regularization term includes the constraints imposed by the recorded data, the boundary conditions
and the initial conditions.

\begin{figure*}[!ht]
\centering
\includegraphics[width=1\columnwidth]{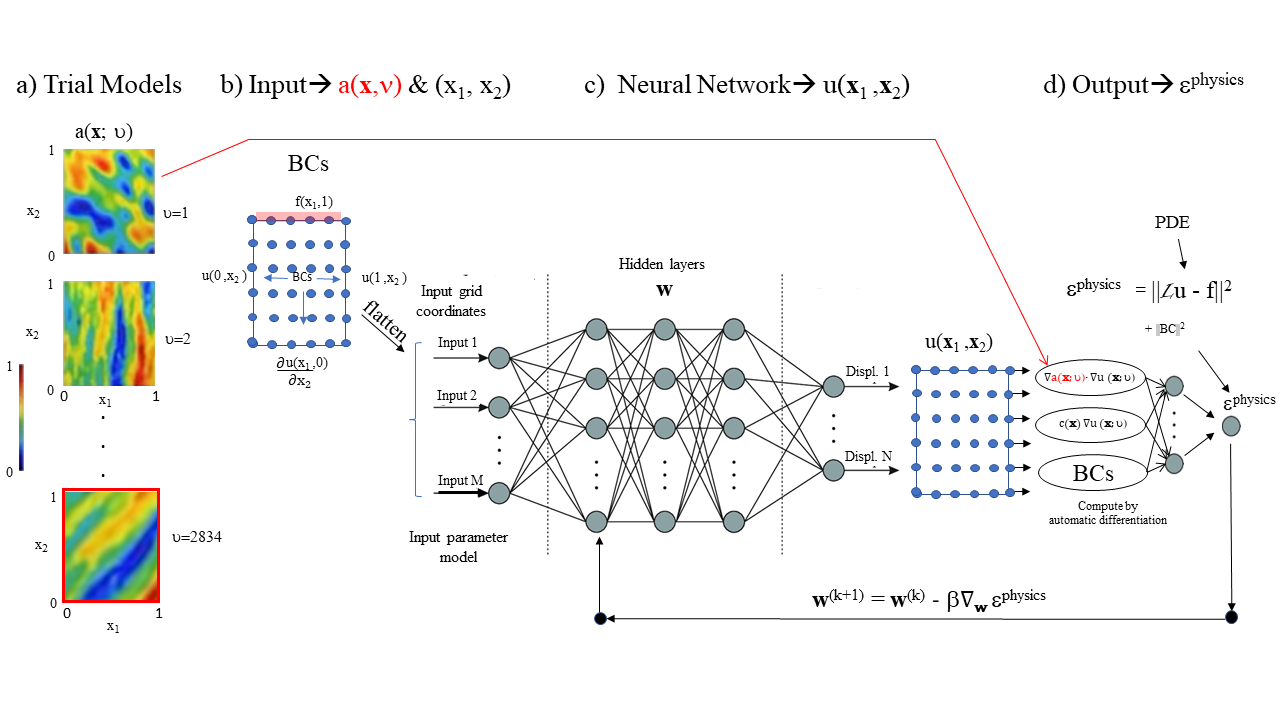}
  \caption{a) Random distribution
  of parameter models, b) input of boundary conditions along the sides
  of the model and the source distribution $f(\x)$ along the top,
  c) the neural network architecture and loss function $\epsilon^{physics}$,
  and d) output field values $u(\x; \nu)$ in the interior of the model are used to form the
  loss function $\epsilon$ as the single scalar output. The gradient $\nabla_{\bf w} \epsilon^{physics}$ can be computed by automatic differentiation of the NN.}
\label{Fig1:PINN.NN}
\end{figure*}

Let us use a simple example to demonstrate the PINN method. Here we use the 2D elliptic equation
\begin{eqnarray}
\overbrace{-\nabla \cdot a(\x;\nu) \nabla u(\x;\nu) + c(\x) u(\x;\nu)}^{\mathcal{L}}
&=& f(\x),\nonumber\\
u=0,~\forall x_1=1;~u=1,~\forall x_1=0;~&&\frac{\partial u}{\partial n} =0,~~ \forall x_2=0,
\label{Inv.FWI.ML.eq3}
\end{eqnarray}
where $\nu$ is a random variable associated with the model parameter function $a(\x;\nu)$.
Figure~\ref{Fig1:PINN.NN}a
depicts  the
computational domain as a 2D unit box where $0\leq x_1\leq 1$, $0\leq x_2\leq 1$,
 where the boundary conditions on the left and right sides are $u(0,x2)=0$ and $u(1,x2)=1$, respectively.
The boundary condition on the bottom of the box at $x_2=0$ is $\partial u/\partial n=0$
and the source function is imposed upon the top of the box at $x_2=1$.
Each selection of $\nu $ leads
to a different parameter model as depicted by the colored images in Figure~\ref{Fig1:PINN.NN}a.

The NN architecture is depicted in Figure~\ref{Fig1:PINN.NN}, where the loss function $\epsilon^{physics}$
is
given by
\begin{eqnarray}
\epsilon^{physics} &=& \overbrace{
<
\sum_{\x_i\in V}
 \{
 -\nabla \cdot a(\x_i;\nu)
\nabla u(\x_i;\nu) + c(\x_i) u(\x_i;\nu) - f(\x_i)
\}^2 >_{\nu}
}^{PDE} \nonumber\\
&&+\overbrace{
<\sum_{\x_i \in \Omega_1} u(\x_i)^2 + \sum_{\x_i\in \Omega_2} \{\frac{\partial u(\x_i)}{\partial n} \}^2>_{\nu}
}^{BC} ,
\label{Inv.FWI.ML.eq3a}
\end{eqnarray}
where the term $<~~>_{\nu}$ indicates averaging over the distribution
of random parameter models as a function of the random variable $\nu$.
Here, $\x_i \in V$ denotes the set of points in the interior of the model, $\x_i \in \Omega_1$ denotes the set of boundary points with Dirichlet boundary conditions, and $\x_i \in \Omega_2$ denotes the set of boundary points
with Neumann boundary conditions.
The NN produces  output field values
throughout the 2D domain in Figure~\ref{Fig1:PINN.NN}c, where the NN is trained on batches of
different input parameter models until the loss function is minimized. The remarkable property here is
that the ground truth field values $u(\x;\nu)$ do not need to be computed by a finite-difference algorithm, they are computed by the
NN. An automatic differentiation
method can be used to compute
their spatial derivatives $\nabla u$ in the loss function of equation~\ref{Inv.FWI.ML.eq3a}; alternatively,
a finite-difference approximation can be applied to the $u(\x;\nu)$ computed by the NN
to get the spatial derivatives.
The weights and biases of the NN are computed by, typically, an Adam gradient
descent method where the residual is back-propagated by automatic differentiation.
Minimizing the loss function ensures that the PDE is satisfied to some specified tolerance.

Minimizing the loss function does not always  guarantee convergence to a global minimum, so
the Dirichlet principle~\citep{karumuri2020simulator} can be used to provide a robust loss function
given by
\begin{eqnarray}
{\mathcal L}(\theta) &=& < \sum_{i \in D} \{ \frac{1}{2} a(\x_i)\nabla \tilde u(\x_i;\nu, \theta)+ c(\x_i )\tilde u^2 (\x_i;\nu,\theta)
-f(\x_i) \tilde u(\x_i;\nu, \theta)  \}
+ \sum_{j\in
\Omega_1+\Omega_2}g_N \tilde u(\x_j;\nu, \theta)>_{\nu} ,\nonumber\\
\label{Inv.FWI.ML.eq4}
\end{eqnarray}
where the last summation is over the boundary points of the parameter model
and $g$ denotes the appropriate operation for the boundary condition.
We now include a tilde over the displacement function $u$ to indicate that it is computed
by the NN with the unknown NN parameters $\theta$.

\begin{figure*}[!ht]
\centering
\includegraphics[width=1\columnwidth]{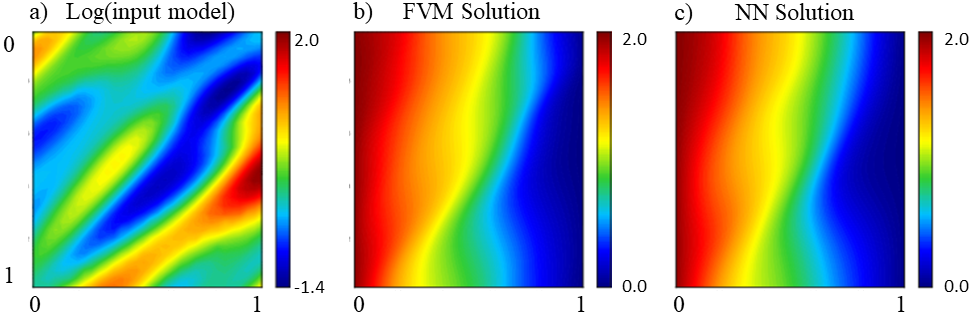}
  \caption{(a) Input parameter model, (b) output $u(x_1,x_2)$ of finite-volume modeling (FVM),
 and  (c) output $u(x_1,x_2)$ of the neural network. Images adapted from~\cite{karumuri2020simulator}.}
\label{Fig1:PINN.NN.f3}
\end{figure*}

\cite{karumuri2020simulator} used equation~\ref{Inv.FWI.ML.eq4}
 to train an NN  by the PINN procedure using several hundred input
parameter models. After training, an input parameter
model is shown in Figure~\ref{Fig1:PINN.NN.f3}a,
and the ground truth displacement field computed by a finite-volume method (FVM)  is depicted in
Figure~\ref{Fig1:PINN.NN.f3}b. The output displacement field computed by the trained NN is shown in
Figure~\ref{Fig1:PINN.NN.f3}c where there are some discrepancies
between it and the ground truth image in Figure~\ref{Fig1:PINN.NN.f3}b.
The low-wavenumber features are accurately computed, but the fine details
of the NN solution are in error. There are at least two causes for these errors: spectral bias and
vanishing gradients.

The spectral bias problem~\citep{rahaman2019spectral} is similar to the multiscale problem in FWI: low-wavenumber features
converge more quickly with iterative FWI, while it takes many more iterations for the solution to converge to the
high-wavenumber features of the model. The partial solution to this problem is to use Fourier features~\citep{rahaman2019spectral,tancik2020fourier},
which transforms the effective neural tangent kernel into a
stationary kernel with a tunable bandwidth.

There is debate as to the advantages of PINNs for forward modeling
compared to the standard modeling methods of finite elements or finite differences.
For modeling in solid mechanics,~\cite{haghighat2021physics} state that their experience
so far suggests that {\it PINN as a forward solver does not lead to advantages in either accuracy or
performance.}

However,~\cite{haghighat2021physics} point out the advantage of PINNs when the
PINN is expensively trained on data and known models, but new data associated with similar
models need to be simulated by PINNs. In this case, the weights of the previously trained
models are used with transfer training as the initial weights for the PINNs, and the new data are efficiently
modeled by PINNs. The authors state that the {\it  advantage of deep learning and PINNs is that a trained PINN is much less costly to execute than classical methods that rely on forward simulations.}
They demonstrate this advantage by making small perturbations of the model parameters, i.e.
changing $\lambda$ and $\mu$ values, to assess the sensitivity
of the data to changes in the model parameters.
Recycling trained weights to act as initial weights
for new models is similar to the frequency-upscaling
approach by~\cite{huang2022pinnup} for forward modeling
of the Helmholtz equation.
Here, they used the NN weights
trained for low-frequency sources
as initial NN weights for modeling the wavefields initiated by high-frequency sources. This frequency-scaling is similar to the multiscale
approach used for FWI~\citep{bunks1995multiscale} where low-pass filtered data
are first inverted for a velocity model, which is then used
as the starting model for data with higher frequencies.

Training of the elastic data is tested with a number of different approaches. In this regard,~\cite{haghighat2021physics} state the following. {\it
 There are three ways to train the network: (1) generate a sufficiently large number of datasets and perform a one-epoch training on each dataset, (2) work on one dataset over many epochs by reshuffling the data, and (3) a combination of these. When dealing with synthetic data, all approaches are feasible to pursue. However, strategy (1) above is usually impossible to apply in practice, especially in space, where sensors are installed at fixed and limited locations. In the original work on PINN~\citep{raissi2019physics}, approach (1) was used to train the model, where datasets are generated on random space discretizations at each epoch. Here, we follow approach (2) to use training data that we could realistically have in practice.}

\subsection{Inverse problem by PINN}\index{Physics-informed neural networks (PINNs)}
The seismic inverse problem in the marine case is defined as finding the velocity model that
is forward modeled to compute the {\it predicted } pressure field $p(\x,t)$ that honors the acoustic
wave equation
\begin{eqnarray}
[\nabla^2 - \frac{1}{c(\x)^2} \frac{\partial^2}{\partial t^2}]p(\x,t) &=&f(\x,t),
\label{Inv.FWI.ML.eq5}
\end{eqnarray}
and agrees with the observed seismograms $p(\x_h,t;\x_s)^{obs}$ recorded along an evenly sampled line
 of hydrophone locations $\x_h$  denoted by
the set of points $\x_h \in {\mathcal D}_{data}$.
We will assume a 2D model and  hydrophones that are fixed either along the ocean floor
or just below the free surface.
Here, $c(\x)$ is the acoustic velocity
in the ocean-earth model and $f(\x_s ,t)$ represents a point source at $\x_s$ excited along an evenly spaced line of points $\x_s \in \mathcal{D}_{src}$, where each source is excited by a bandlimited
wavelet $w(t)$. The hydrophone line records the pressure fields excited by  each source
and is below and parallel to the line of source locations at $\x_{s} \in {\mathcal D}_{src}$. The boundary condition on the ocean's surface is $p(x_1,x_2=0,t)=0~~\forall x_1 $ and all-time samples $0\leq t \leq T$. The bottom and sides of the model
have absorbing boundary conditions~\citep{yilmaz2001seismic}.

Agreement between the observed $p(\x_i,t_k;\x_j)^{obs}$
and NN predicted $p(\x_i,t_k;\x_j)$ seismograms is defined as the velocity model that minimizes the
sum of the squared residuals:
\begin{eqnarray}
\epsilon^{data}&=&\frac{1}{M_D}
\sum_{\x_i \in {\mathcal D}_{data}}
\sum_{\x_j \in {\mathcal D}_{src}}
\sum_{t_k \in {\mathcal D}_{time}}
(p(\x_i,t_k;\x_j)-p(\x_i,t_k;\x_j)^{obs})^2,
\label{Inv.FWI.ML.eq6}
\end{eqnarray}
where ${\mathcal D}_{data}$ represents the set of
hydrophone locations just below the free surface, ${\mathcal D}_{time}$ is the set of time samples,
and $M_D$ is the total number of sampled points in the summand.
The objective function for computing the pressure field that satisfies
the acoustic wave equation by an NN is defined as
\begin{eqnarray}
\mathcal{\epsilon}^{physics}&=&
\frac{1}{M_p}
\sum_{\x_j \in {\mathcal D}_{src}}
\sum_{\x_l \in {\mathcal D}_{model}}
\sum_{t_k \in {\mathcal D}_{time}}
\{\overbrace{[\nabla^2 - \frac{1}{c(\x_l)^2} \frac{\partial^2}{\partial t^2}]}^{{\mathcal L}} p(\x_l,t_k;\x_j)
 -f(\x_j,t_k)\}^2;\nonumber\\
\label{Inv.FWI.ML.eq7}
\end{eqnarray}
where ${\mathcal D}_{model}$ is the set of  sampled points in the velocity model
and $M_p$ is the total number of sampled points in the summand.

The boundary conditions on the free surface and the absorbing boundary conditions,
are included in
\begin{eqnarray}
 \epsilon^{BC} &=&\frac{1}{M_{BC}}
\sum_{i \in {\mathcal D}_{boundary}} |A_i|^2,
 \label{Inv.FWI.ML.eq7a}
\end{eqnarray}
where $A_i$ represents the boundary conditions associated with
all the points on the boundary and all the time samples, ${\mathcal D}_{boundary}$
represents the set of indices associated with these samples, and $M_{BC}$
is the total number of boundary samples in both time and space.
For example, the boundary conditions might take on the following form:
\begin{eqnarray}
Left~ABC:&&~~\frac{\partial p(\x,t)}{\partial t}-\frac{1}{c(\x)} \frac{\partial p(\x,t)}{\partial x_1}=0, ~\x=(x_1=0,x_2)~for~ 0\leq x_2\leq 1,\nonumber\\
Right~ABC:&&~~\frac{\partial p(\x,t)}{\partial t}+\frac{1}{c(\x)} \frac{\partial p(\x,t)}{\partial x_1}=0, ~\x=(x_1=1,x_2)~for ~0\leq x_2\leq 1,\nonumber\\
Bottom~ABC:&&~~\frac{\partial p(\x,t)}{\partial t}+\frac{1}{c(\x)}\frac{\partial p(\x,t)}{\partial x_2}=0, ~\x=(x_1,x_2=1)~for~ 0\leq x_1\leq 1,\nonumber\\
Free~Surface:&&~~p(\x,t)=0, ~\x=(x_1,x_2=0)~~for ~0\leq x_1\leq 1,
\label{Inv.FWI.ML.eq8}
\end{eqnarray}
where $c$ takes on the velocity near the boundary point of interest.
The source wavelet $w(t)$
turns on shortly after $t=0$ to excite the model.
The initial conditions can be enforced by setting $p(\x,t)$ and $\frac{\partial p(\x,t)}{\partial t}=0$ at $t=0$ $\forall \x$ with the loss function
denoted as $\epsilon^{IC}$.

\begin{figure*}[!ht]
\centering
\includegraphics[width=1\columnwidth]{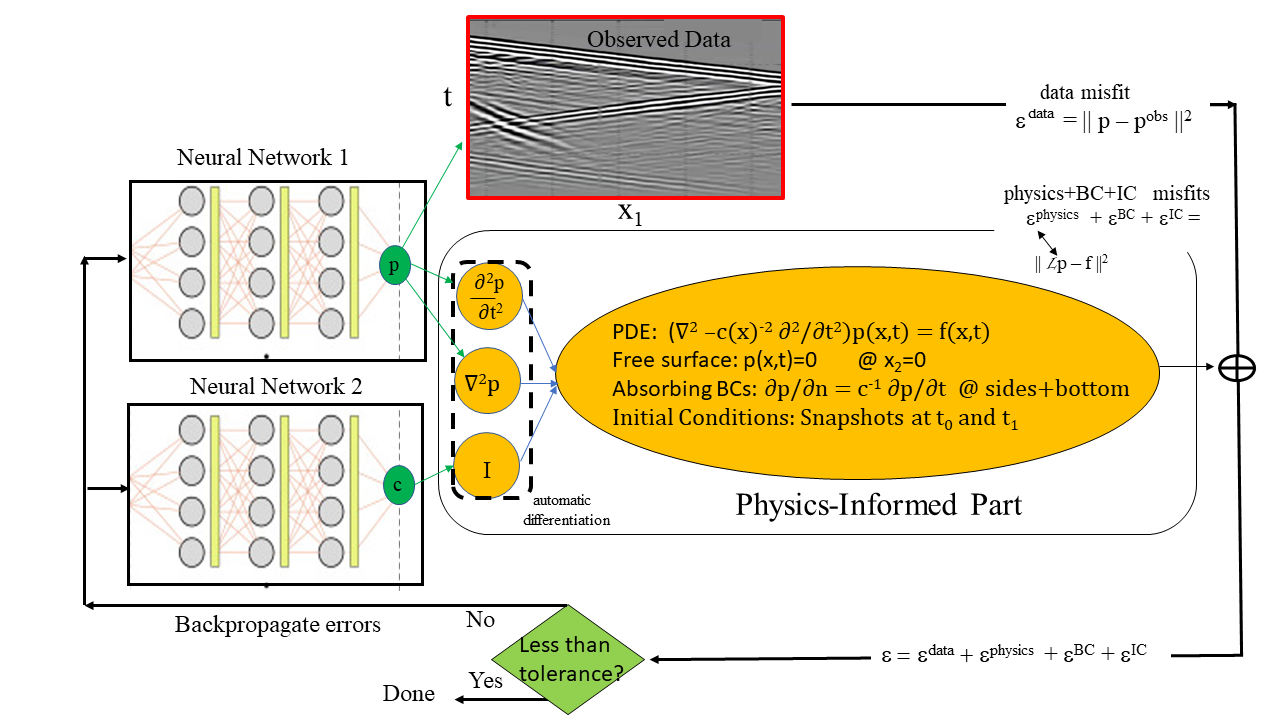}
\caption{Workflow for inversion of seismic data by PINN
using the objective function in equation~\ref{Inv.FWI.ML.eq10}. A pass through the data loss term is
carried out with a batch of data, and then its output is used to carry out a pass through the physics loss term.
Adapted from~\cite{rasht2022physics}.}
\label{Fig1:PINN.NN.f20}
\end{figure*}

Therefore, the inverse PINN objective function $\epsilon$ is the
combination of the objective functions in
equations~\ref{Inv.FWI.ML.eq6}-\ref{Inv.FWI.ML.eq7a} and $\epsilon^{IC}$:
\begin{eqnarray}
\epsilon^{PINNi} &=& \epsilon^{data} + \epsilon^{physics} + \epsilon^{BC}+\epsilon^{IC}+
regularization,
\label{Inv.FWI.ML.eq10}
\end{eqnarray}
where the schematic description is in Figure~\ref{Fig1:PINN.NN.f20}.

The spatial and temporal grids do not have to be evenly sampled, and are often sampled at random points both in space and time.
This suggests that some cost savings can be made by reducing the number of points at which the
PDE must be solved.
However, the low-frequency details appear to lead to faster convergence
than the higher-frequency details. This suggests that
a multiscale approach should be useful, where low-frequency
data is first inverted, and the low-wavenumber model and PINN weights (via transfer learning)
are used as the starting model
for the next higher-frequency data. Another possibility is to use the low-wavenumber inverted model
as the starting model for standard FWI~\citep{song2021wavefield}.~\cite{muller2022deep} use transfer training from low-wavenumber models and simple data
to capture the main velocity trend of the actual model. They used PINN to
 stabilize the inversion, acting like a regularizer and avoiding local minima-related problems.

\bibstyle{seg} 
\bibliography{main}


\end{document}